\documentclass[%
 aip,
amsmath,amssymb,
 reprint,%
]{revtex4-1}

\usepackage{graphicx}
\usepackage{dcolumn}
\usepackage{bm}

\usepackage[utf8]{inputenc}
\usepackage[T1]{fontenc}
\usepackage{mathptmx}
\usepackage{etoolbox}
\usepackage[usenames,dvipsnames]{xcolor}
\usepackage{mciteplus}

\makeatletter
\def\@email#1#2{%
 \endgroup
 \patchcmd{\titleblock@produce}
  {\frontmatter@RRAPformat}
  {\frontmatter@RRAPformat{\produce@RRAP{*#1\href{mailto:#2}{#2}}}\frontmatter@RRAPformat}
  {}{}
}%
\makeatother

\begin{document}

\title{Thermalized Abrikosov lattices from decaying turbulence in rotating BECs}
\author{Julian Amette Estrada}
\email{julianamette@df.uba.ar}
\affiliation{Universidad de Buenos Aires, Facultad de Ciencias Exactas y Naturales, Departamento de Física. Ciudad Universitaria, 1428 Buenos Aires, Argentina.}
\affiliation{CONICET - Universidad de Buenos Aires, Instituto de F\'{\i}sica del Plasma (INFIP). Ciudad Universitaria, 1428 Buenos Aires, Argentina.}

\author{Marc E.~Brachet}
\email{brachet@phys.ens.fr}
\affiliation{Laboratoire de Physique de l'{E}cole {N}ormale {S}up\'erieure, ENS, Universit\'e PSL, CNRS, Sorbonne Universit\'e, Universit\'e de Paris, F-75005 Paris, France}

\author{Pablo D.~Mininni}
\email{mininni@df.uba.ar}
\affiliation{Universidad de Buenos Aires, Facultad de Ciencias Exactas y Naturales, Departamento de Física. Ciudad Universitaria, 1428 Buenos Aires, Argentina.}
\affiliation{CONICET - Universidad de Buenos Aires, Instituto de F\'{\i}sica del Plasma (INFIP). Ciudad Universitaria, 1428 Buenos Aires, Argentina.}

%

\date{\today}

\begin{abstract}
We study the long-time decay of rotating turbulence in Bose-Einstein condensates (BECs). We consider the Gross-Pitaevskii equation in a rotating frame of reference, and review different formulations for the Hamiltonian of a rotating BEC. We discuss how the energy can be decomposed, and present a method to generate out-of-equilibrium initial conditions. We also present a method to generate finite-temperature states of rotating BECs compatible with the Canonical or the Grand canonical ensembles. Finally, we integrate numerically rotating BECs in cigar-shaped traps. A transition is found in the system dynamics as the rotation rate is increased, with a final state of the decay of the turbulent flow compatible with an Abrikosov lattice in a finite-temperature thermalized state.
\end{abstract}

\maketitle

\section{Introduction}

When a quantum fluid is perturbed out of equilibrium with a velocity above a certain critical threshold, the system spontaneously develops topological defects which correspond to quantum vortices. These vortices can interact nonlinearly, giving rise to the phenomenon of quantum turbulence. This is the case in out-of-equilibrium Bose-Einstein condensates (BECs)\cite{Henn2009, Tsatsos_2016}, as well as in superfluid $^4$He and $^3$He-B. However, turbulence is classically expected to dissipate energy efficiently, returning the system to a new equilibrium. In the quantum case it is unclear how these systems relax back to an equilibrium, and what are the properties of the resulting state.

Turbulence is ubiquitous in nature, and BECs and superfluids are no exception. There are several laboratory experiments of quantum turbulence \cite{Coddington2003, Bewley2006, Henn2009, White2014, Tsatsos_2016, makinen2022}, as well as theoretical and numerical studies \cite{Nore1997, Lvov2010, Laurie_2010, ClarkdiLeoni2015, Shukla2019, Mller2020, Hossain2022}. In most cases, quantum turbulence resembles its classical counterpart. Finding differences between them has become a key part to improve our understanding of both systems, as well as of out-of-equilibrium systems in general. In many cases both quantum and classical turbulence display Kolmogorov scaling $E(k) \sim k^{-5/3}$ of the incompressible kinetic energy spectrum, but the mechanisms behind this behaviour are believed to be drastically different. At large scales vortex reconnection in quantum turbulence transfers energy to smaller scales, just as in classical turbulence. But for scales smaller than the intervortex distance a cascade of Kelvin waves, negligible in most cases in classical turbulence, is responsible for the transfer of energy towards smaller scales, where it can be dissipated through phonon emission \cite{Lvov2010}. Some experiments in quantum turbulence \cite{Vinen1957, Walmsley2008, Barenghi2014, Walmsley2012, Walmsley2014} observe a different scaling known as Vinen turbulence (or ``ultraquantum'' regime). This regime has no classical counterpart and is characterized by the decay of the total vortex length as $\sim t^{-1}$, compatible with a scaling of the isotropic spectrum of the incompressible kinetic energy as $E(k) \sim k^{-1}$. In this regime, the presence of a non-condensed phase is believed to play a crucial role. This behavior was indeed also reported in simulations with counterflows \cite{Baggaley2012, Cidrim2017}. Nonetheless, more recent simulations of BECs without a counterflow found the same spectral scaling for an initial array of ordered vortices \cite{Cidrim2017, Marino2021}, in homogeneous superfluid turbulence \cite{Polanco_2021}, and in freely decaying rotating quantum turbulence in BECs \cite{AmetteEstrada2022}.

Although studies of turbulence in rotating BECs and superfluids are very recent, equilibrium and near-equilibrium properties of rotating condensates have been studied in detail. When rotation is present the quantum nature of the condensate results in a transition from a state without vortices into one with ordered vortex lattices \cite{Fetter_2001, Cooper_2004, Fetter_2009}. In this regime, the condensate displays global oscillation modes and waves that do not have a classical counterpart \cite{Tkachenko1965, Andereck1982, Sonin_2005}. For even larger values of the rotation rate the system displays solutions that link its properties with other problems in condensed matter physics, such as Landau levels in magnetic systems or the quantum Hall effect \cite{Fetter_2009}. Rotating quantum turbulence thus becomes an interesting problem in which the flow perturbs the system out of an equilibrium with no classical counterpart \cite{Walmsley2012, AmetteEstrada2022, Hossain2022}. Recent numerical studies considered rotating turbulence in BECs \cite{AmetteEstrada2022} and in unitary Fermi gases \cite{Hossain2022}. While the former study found differences in the dissipation mechanism and phonon emission with the non-rotating case, the latter reported differences in the dissipation mechanisms between fermionic and bosonic superfluids. A detailed comparison of the many possible regimes in this system with classical turbulence is still missing. In classical turbulence, rotation induces two important changes in the dynamics: the flow becomes quasi-two-dimensional (2D), and  a steeper-than-Kolmogorov spectrum $E(k) \sim k^{-2}$ develops at small scales \cite{Waleffe_1993, Cambon_1997, Cambon_2004, Pouquet_2010}. Inertial waves play a central role in setting this spectrum and the energy cascade rate, while in finite domains the flow at large scales also self-organizes into columnar vortices with an inverse cascade of energy \cite{Sen_2012, Clark_Di_Leoni_2020b}.

To understand how these flows relax back to an equilibrium as turbulence decays, and the properties of the final states, a method to generate equilibria at finite temperature is needed. BECs at zero temperature can be modeled by the Gross-Pitaevskii equation, which is a non-perturbative mean-field equation for a classical field \cite{Davis2001, Proukakis2008}. Many extensions of this equation to consider systems at finite temperature exist \cite{Gardiner_2002, Calzetta_2007, Proukakis2008, Krstulovic2011,Berloff2014}, in most cases involving some truncation or coarse-graining of the system, together with a stochastic formulation of the equation for the classical field. In the case of non-rotating quantum turbulence, similar approaches have been used to study properties of the thermalized system as the flow decayed \cite{Krstulovic2011, Shukla2019}.

In this work we consider freely decaying turbulence in rotating BECs, and study its decay for long times. We consider the Gross-Pitaevskii equation in a rotating frame of reference, write the Madelung transformation that allows computation of the fluid velocity, and review different formulations for the Hamiltonian of a rotating BEC. We also discuss how the total system energy can be decomposed in the rotating case, as well as methods to generate out-of-equilibrium initial conditions. We present a method to generate finite-temperature thermalized states of rotating BECs compatible with Canonical or Grand canonical equilibria. Then, we integrate numerically turbulent rotating condensates in cigar-shaped traps. We show there is a transition in the system dynamics as the rotation rate increases above a critical value, and we show that the final state of the decay of the turbulent flow is compatible with an Abrikosov lattice in a thermalized state with finite temperature.

\section{Rotating Bose-Einstein condensates} 
\label{sec:methods}
\subsection{The rotating Gross-Pitaevskii equation}

We describe BECs using the Gross-Pitaevskii equation (GPE) in a rotating frame of reference, with angular rotation $\bm{\Omega} = \Omega \hat{z}$ in a trapping harmonic potential $V(\bm{x})$. In the following we call this equation the rotating Gross-Pitaevskii equation (RGPE). It describes the evolution of a condensate of weakly interacting bosons of mass $m$ under the aforementioned conditions, and is expressed as,
\begin{equation}
     i \hbar \frac{\partial \psi (\bm{r},t)}{\partial t} = \left[  -\frac{\hbar^2 \nabla^2}{2 m} + g |\psi (\bm{r},t)|^2 + V(\bm{r}) - \Omega J_z \right]\psi (\bm{r},t),
   \label{eq:RGPE}
\end{equation}
where $g$ is related to the scattering length, $J_z = \hat{z}\cdot (\bm{r} \times \bm{p})$ is the angular momentum operator along the axis of rotation, and $\bm{p} = -i \hbar \boldsymbol{\nabla}$ is the momentum operator. This equation can be derived by applying the time-dependent constant-speed rotation operator $R(t, \boldsymbol{\Omega})$ to the GPE, and by defining the rotated order parameter (or ``wave function'') as $\psi(\bm{r},t) = R(t,\boldsymbol{\Omega}) \psi(\bm{r}',t)$, where $\psi(\bm{r}',t)$ is the order parameter in the non-rotated laboratory coordinate system.

In the $\Omega = 0$ case the RGPE (or equivalently in this case, the GPE in the non-rotating frame) can be mapped to the Euler equation for an isentropic, compressible, and irrotational fluid  with an extra pressure term due to quantum effects (known as the quantum pressure). This is done by means of the Madelung transformation \cite{Nore1997}
\begin{equation}
    \psi (\bm{r}',t) = \sqrt{\rho(\bm{r}',t)/m} \, e^{i S(\bm{r}',t)},
\end{equation}
where $\rho(\bm{r}',t)$ is the fluid mass density and $S(\bm{r}',t)$ is the phase of the order parameter. With this choice, the fluid velocity is $\bm{v}' = \bm{v}(\bm{r}',t) = (\hbar/m) \bm{\nabla}' S(\bm{r}',t)$. The resulting flow is thus irrotational except for topological defects where the vorticity is quantized so that $\oint_{\mathcal{C}} \bm{v}' \cdot d \bm{l}' = \Gamma_0 n$ with $n \in \mathbb{N}$, and where $\Gamma_0 = h / m$ is the quantum of circulation.

When $\Omega \neq 0$ an analogous mapping can be used in the rotating frame. The rotating Euler equation (with the extra quantum pressure term, but this time also with Coriolis and centrifugal forces) is obtained by defining the velocity in the rotating frame as $\bm{v}_\textrm{rot}^{\prime} = (\hbar/m) \bm{\nabla}' S(\bm{r}',t)- \boldsymbol{\Omega} \times {\bm r}' = \bm{v}' - \boldsymbol{\Omega} \times {\bm r}'$, where $\bm{v}'$ is the velocity in the non-rotating frame. This is consistent with a classical rotation, and for $\Omega = 0$ we recover the previous case. However, the velocity in the rotating frame cannot be obtained directly as the gradient of the phase in a Madelung transformation applied to $\psi(\bm{r},t)$. Indeed, if a Madelung transformation of the form 
$\psi(\bm{r},t) = \sqrt{\rho(\bm{r},t)/m} \, e^{iS(\bm{r},t)}$ is used, and a continuity equation $\partial \rho/\partial t + \boldsymbol{\nabla} \cdot \bm{j}=0$ is derived from Eq.~\eqref{eq:RGPE}, it follows that
\begin{equation}
\bm{j}(\bm{r},t) = -\frac{i \hbar}{2} (\psi^* \boldsymbol{\nabla} \psi - \psi \boldsymbol{\nabla} \psi^*) - \rho \boldsymbol{\Omega} \times \bm{r} .
\label{eq:current}
\end{equation}
Thus, associating $\bm{j}(\bm{r},t) = \rho \bm{v}_\textrm{rot}$ we have again $\bm{v}_\textrm{rot} = (\hbar/m) \bm{\nabla} S(\bm{r},t)- \boldsymbol{\Omega} \times {\bm r} = \bm{v} - \boldsymbol{\Omega} \times {\bm r}$, now in terms of the $\bm{r}$ coordinate. In other words, $\rho$ and $\rho'$, and $\bm{v}$ and $\bm{v}'$, are the same fields just passively rotated. Note also that all these relations follow from the fact that the velocity of the superfluid must be irrotational everywhere except for topological defects, and thus the solid body rotation can only be mimicked by the quantum flow by generating an array of quantum vortices. We will discuss this next in the context of the Hamiltonian of these systems.

The Hamiltonian $\mathcal{H}_0^{\prime}$ corresponding to a condensate of weakly interacting bosons described by GPE (i.e., in the non-rotating case) is
\begin{multline}
\mathcal{H}_0^{\prime}[\psi',\psi'^*] = \int d^3r' \left[ \frac{\hbar^2}{2m} |\boldsymbol{\nabla}' \psi'|^2 + \frac{g}{2} |\psi'|^4 + V(\bm{r}') |\psi'|^2 \right],
\end{multline}
where the asterisk denotes the complex conjugate. Using the Madelung transformation and the relations $|\psi'|^2=\rho'/m$ and $\hbar^2 |\boldsymbol{\nabla}' \psi'|^2/(2m) = \rho' |\bm{v}'|^2/2 + \hbar^2 |\boldsymbol{\nabla}' \sqrt{\rho'}|^2/(2m^2)$, we can decompose the energy of our system into fluid-like energy components as\cite{Nore1997}
\begin{align}
    E_k &= \frac{1}{2} \langle \rho' |\bm{v}'|^2 \rangle , \\
    E_q &= \frac{\hbar^2}{2m^2} \langle |\boldsymbol{\nabla}' \sqrt{\rho'}|^2 \rangle , \\
    E_p &= \frac{g}{2m^2} \langle \rho'^2 \rangle , \\
    E_V &= \frac{1}{m} \langle V' \rho' \rangle ,
\end{align}
where the brackets denote volume average (note that, strictly speaking, these are mean energy densities). Here $E_k$ is the flow kinetic energy, $E_q$ the quantum energy, $E_p$ the internal (or potential) energy, and $E_V$ is potential energy associated to the trap. By means of the Helmholtz decomposition we can further decompose\cite{Nore1997} $(\sqrt{\rho'} \bm{v}')=(\sqrt{\rho'} \bm{v}')^{\rm (c)}+ (\sqrt{\rho'} \bm{v}')^{\rm (i)}$, where the superindices c and i denote respectively the compressible (irrotational) and incompressible (solenoidal) vector field components (i.e., such that $\boldsymbol{\nabla} \times (\sqrt{\rho'} \bm{v}')^{\rm (c)}=0$ and $\boldsymbol{\nabla} \cdot (\sqrt{\rho'} \bm{v}')^{\rm (i)}=0$ respectively). With this decomposition, the kinetic energy $E_k$ can be further separated into the compressible $E_k^{\rm (c)}$ and incompressible $E_k^{\rm (i)}$ kinetic energies, as often done in the study of classical compressible flows \cite{KidaOrszag1990}. With these definitions, Parseval's identity allows us to define wave number power spectra for all energies \cite{Nore1997, ClarkdiLeoni2015}. 

Rotation changes the Hamiltonian, adding in the rotating frame a term
\begin{equation}
\mathcal{H}_{rot} = - \bm{\Omega} \cdot \int d^3r \, \psi^* (\bm{r}\times\bm{p}) \psi .
\end{equation}
This term lets us define a new energy component, the rotational energy, as
\begin{equation}
    E_{rot} = - \bm{\Omega}  \cdot \langle \psi^* \bm{J} \psi \rangle.
\end{equation}
The total Hamiltonian for the condensate in the rotating frame is then $\mathcal{H} = \mathcal{H}_0[\psi,\psi^*] + \mathcal{H}_{rot}$. When rotation is present, the kinetic part of the Hamiltonian may be rewritten using the Madelung transformation, resulting in
\begin{eqnarray}
    \mathcal{H} &=& \int d^3r \Bigg[ \frac{\rho}{2} |\bm{v}|^2 + \frac{\hbar^2}{2m^2} |\boldsymbol{\nabla} \sqrt{\rho}|^2 + \frac{g}{2} |\psi|^4  \nonumber \\
    {} &&
    + V(\bm{r}) |\psi|^2 - \boldsymbol{\Omega} \cdot (\psi^* \bm{J} \psi) \Bigg] .
\end{eqnarray}
From the discussion before, it follows that $E_k=\langle \rho' |\bm{v}'|^2 \rangle/2 = \langle \rho |\bm{v}|^2 \rangle/2$, and similar relations hold for all integral and volume averages of hydrodynamic scalar and vector fields. Thus, from this Hamiltonian we can now decompose the energy as $E = E_k + E_q + E_p + E_V + E_{rot}$. Using the relation\cite{Fetter_2001} $-\boldsymbol{\Omega} \cdot (\psi^* \bm{J} \psi) = i \hbar/(2 m) \boldsymbol{\nabla} \cdot (\rho \boldsymbol{\Omega} \times \bm{r}) - \rho \bm{v} \cdot (\boldsymbol{\Omega} \times \bm{r})$ we can also write this Hamiltonian as
\begin{eqnarray}
    \mathcal{H} &=& \int d^3r \Bigg[ \frac{\rho}{2} (\bm{v} - \boldsymbol{\Omega} \times \bm{r})^2 + \frac{\hbar^2}{2m^2} |\boldsymbol{\nabla} \sqrt{\rho}|^2 \nonumber \\
    {} &&
    + \frac{g}{2m^2} \rho^2 + \left( \frac{V(\bm{r})}{m} - \frac{(\bm{\Omega} \times \bm{r})^2}{2} \right) \rho \Bigg] .
    \label{eq: rotating hamiltonian}
\end{eqnarray}
The first term is associated to the kinetic energy density in the rotating frame, $\langle \rho |\bm{v}_\textrm{rot}|^2 \rangle/2$, while the last term is a repulsive centrifugal correction to the trapping potential.

In order to minimize the energy, the velocity $\bm{v}$ in this expression must equal the rigid body rotation. This corresponds to the the case in which $\bm{v}_\textrm{rot}$ is minimum. But while this is what would happen in a normal fluid, in a superfluid $\bm{v}$ must be irrotational everywhere except for the topological defects. Thus, the system can only mimic $\boldsymbol{\Omega} \times \bm{r}$ by generating quantum vortices. Moreover, the system cannot generate less than one quantum of circulation. As a result, for this process to happen the rotation rate must be above the threshold\cite{Fetter_2009} $\Omega_c = 5\hbar /(2 m R_\perp^2) \ln (R_\perp/\xi)$ (where $R_\perp$ is the condensate radius, and $\xi$ the condensate healing length) such that creating one quantized vortex results in a state with less free energy than having none. For larger values of $\Omega$, rotation forces the system into a two-dimensional (2D) state that mimics the solid body rotation via the generation of a regular array of quantum vortices. This array is known as the Abrikosov lattice, and is such that its total circulation approximates that of the rotation. To this end it must have a density of vortices per unit area of $n_v = \Omega / (\sqrt{2}\pi c \xi)$, where $c$ is the speed of sound. Tkachenko\cite{Tkachenko1965} showed that for an infinite homogeneous system ($V(\bm{r})=0$) this lattice must be triangular to minimize the free energy of the system. 

When such an equilibrium is perturbed, the system can sustain waves (sound or phonons, Kelvin waves, inertial waves, and Tkachenko waves which correspond to normal modes of the Abrikosov lattice). For details of the dispersion relations for each of these waves, see Ref.~\cite{AmetteEstrada2022}. Abrikosov lattices and Tkachenko waves were experimentally observed in rotating BECs\cite{Coddington2003}. Similar lattices were observed in superfluid helium \cite{Yarmchuk79}. Vortex lattices were also reported to be metastable states of classical rotating turbulence in finite domains \cite{Clark_Di_Leoni_2020b}. Finally, it's also worth pointing out the similarity between the system under study and type II superconductors in external magnetic fields. As an example, taking the curl of Eq.~\eqref{eq:current} gives 
\begin{equation}
    \bm{\nabla} \times \bm{j} = -2 \rho \bm{\Omega},
\end{equation}
which is equivalent to one of London equations in superconductors. The similarity between the rotation and a magnetic field can also be seen in the system Hamiltonian.

\subsection{Generation of initial conditions for RGPE}

We now discuss how to generate initial conditions for RGPE compatible with the system rotation, either to have steady Abrikosov lattices, or to generate flows that can evolve into turbulent states at very low temperature. To that end, we need a method to generate low acoustic emission states that are quasi-stationary, and that can be used as initial conditions for Eq.~\eqref{eq:RGPE}. 

Given some time independent solution of RGPE, one can verify that it must also be a solution of the dissipative rotating real Ginzburg-Landau equation (RRGLE),
\begin{equation}
    \frac{\partial \psi}{\partial t} = \left(  \frac{\hbar}{2 m}\nabla^2 - \frac{g}{\hbar} |\psi |^2 + \frac{V}{\hbar} - \frac{\Omega J_z}{\hbar} + \frac{\mu}{\hbar} \right) \psi ,
    \label{eq: RGLE}
\end{equation}
where $\mu$ is the chemical potential. Note that the extra term proportional to $\mu$ can be also added into RGPE, but it does not affect the global dynamics of the system as it only adds a global phase to the order parameter, turning $\psi$ into $e^{- i \mu t/\hbar} \psi$.  

Equation \eqref{eq: RGLE} suffices to prepare, e.g., a condensate with an Abrikosov lattice. The equation can be integrated numerically, and the solution will decay into a condensate trapped by the harmonic potential, while a lattice of vortices will appear to adjust for the system rotation. To generate initial conditions for RGPE with a non-zero flow in the rotating frame of reference, that can be used to study the development and evolution of turbulence, one wants to perturb the system with a non-trivial velocity field $\bm{u}$. To do this, following the method in Ref. \cite{Nore1997}, one can use the asymptotic solutions of the rotating-advective real Ginzburg-Landau equation (RARGLE)
\begin{equation}
     \frac{\partial \psi}{\partial t} = \left( \frac{\hbar}{2 m} \nabla^2 - \frac{g}{\hbar} |\psi|^2 - \frac{V}{\hbar} + \frac{\Omega J_z}{\hbar} + \frac{\mu}{\hbar} - i {\bm{u}} \cdot \boldsymbol{\nabla} - \frac{m|\bm {u}|^2}{2\hbar} \right) \psi.
   \label{eq:RARGLE}
\end{equation}
Solutions to this equation have the desired properties of low acoustic emission and quasi-stationarity. Besides, the resulting order parameter (used as an initial condition for RGPE) generates a velocity field $\bm{u}$. This can be seen explicit by writing the energy that the time asymptotic solutions of Eq.~\eqref{eq:RARGLE} minimize, which is given by,
\begin{eqnarray}
    \mathcal{H}_{RARGL} &=& \int d^3r \Bigg[ \frac{\rho}{2} (\bm{v}-\bm{u})^2 + \frac{\hbar^2}{2m^2} |\boldsymbol{\nabla} \sqrt{\rho}|^2 + \frac{g}{2} |\psi|^4  \nonumber \\
    {} &&
    + V(\bm{r}) |\psi|^2 - \boldsymbol{\Omega} \cdot (\psi^* \bm{J} \psi) \Bigg] .
\end{eqnarray}
Thus, Eq.~\eqref{eq:RARGLE} generates states in which the superfluid velocity $\bm{v}$ tries to approximate the advection velocity $\bm{u}$ while also generating a rotation with angular velocity $\boldsymbol{\Omega}$.

\subsection{Finite temperature model and statistical ensembles}

If one considers the condensate as part of a larger system $\mathcal{S}$ with whom it can exchange both particles and energy, equilibria will be characterized by fixed values of the system volume $\mathcal{V}$, the chemical potential, and the temperature $T$. The probability is then given by the Grand canonical ensemble,
\begin{equation}
    \mathbb{P} = \frac{e^{-\beta (\mathcal{H} - \mu \mathcal{N})}}{\mathcal{Z}},
\end{equation}
where $\beta=1/(k_B T)$, $k_B$ is the Boltzmann constant, $\mathcal{Z}$ is the Grand canonical partition function, and $\mathcal{N}$ is the number of particles in the system.

These equilibrium states are difficult to find, as the Hamiltonian has terms that are quartic in the order parameter. However, the method presented in Ref.~\cite{Krstulovic2011} can be generalized to systems in rotating frames. First, a forcing with Gaussian-white noise is added to Eq.~\eqref{eq: RGLE}, defining the rotating Ginzburg-Landau equation at finite temperature (RGLET), 
\begin{eqnarray}
    \frac{\partial \psi}{\partial t} &=& \left[  \frac{\hbar}{2 m}\nabla^2 - \frac{g}{\hbar} |\psi |^2 + V(\bm{r}) - \frac{\Omega Jz}{\hbar} + \frac{\mu}{\hbar} \right]\psi + \nonumber \\ 
    {} &&
    \sqrt{\frac{2 }{\mathcal{V} \hbar \beta}} \zeta (\bm{r},t) ,
    \label{eq: RGLET}
\end{eqnarray}
where $\zeta (\bm{r},t)$ is a delta-correlated random process such that $\left< \zeta (\bm{r},t) \zeta^* (\bm{r}',t')\right> = \delta (\bm{r} - \bm{r}') \delta (t-t')$, and the factor $\sqrt{2 / \mathcal{V} \hbar \beta}$ controls the amplitude of the fluctuations through the temperature $T$.

Next, we perform a Galerkin truncation of Eq.~\eqref{eq: RGLET} in Fourier space. That is, the expansion of $\psi$ and Eq.~\eqref{eq: RGLET} are truncated using the Galerkin projection operator defined as
\begin{equation}
    \mathcal{P}_G(\hat{f}(\bm{k})) = \Theta(k_{\textrm{max}}-|\bm{k}|) \hat{f}(\bm{k}),
\end{equation}
where $\hat{f}(\bm{k})$ is the Fourier transform of $f(\bm{r})$, $k_{\textrm{max}}$ is some maximum (cutoff) wave number (physically $k_{\textrm{max}}>2\pi /\xi$, in practice in the numerical simulations $k_{\textrm{max}}$ is the maximum wave number that the simulations can resolve), and $\Theta$ is the Heaviside function. To conserve the momentum, the term proportional to $|\psi|^2 \psi$ in Eq.~\eqref{eq: RGLET} must be truncated as\cite{Krstulovic2011} $\mathcal{P}_G(\mathcal{P}_G(|\psi|^2) \psi)$. From now on we simplify the notation by doing $\mathcal{P}_G(\hat{f}(\bm{k})) \rightarrow \hat{f}(\bm{k})$. Defining the free energy $F = \mathcal{H} - \mu N$, the truncated Eq.~\eqref{eq: RGLET} can be written as a Langevin equation for the evolution of each Fourier mode of $\psi$,
\begin{equation}
    \frac{\partial \hat{\psi}(\bm{k},t)}{\partial t} = - \frac{1}{\mathcal{V} \hbar} \frac{\partial F }{\partial \hat{\psi}^* (\bm{k},t)} + \sqrt{\frac{2}{\mathcal{V} \hbar \beta}} \hat{\zeta}(\bm{k}, t) ,
    \label{eq: langevin fourier}
\end{equation}
where the functional $F$ is $F\left[\{\hat{\psi} (\bm{k},t), \hat{\psi}^* (\bm{k},t)\right\}]$ (for all $|\bm{k}| < k_{\textrm{max}}$). The resulting stochastic process has a state probability $\mathbb{P}[\{\hat{\psi} (\bm{k},t), \hat{\psi}^* (\bm{k},t)\}]$ that is described by a corresponding multivariate Fokker-Planck equation \cite{VanKampen,Krstulovic2011}
\begin{equation}
    \frac{\partial \mathbb{P}}{\partial t} = \sum_{\bm{|k| < k_{\textrm{max}}}} \frac{\partial}{\partial \hat{\psi}_{\bm{k}}} \left[ \frac{1}{\mathcal{V} \hbar} \frac{\partial F}{\partial \hat{\psi}_{\bm{k}}^*} \mathbb{P} + \frac{1}{\mathcal{V} \hbar \beta} \frac{\partial \mathbb{P}}{\partial \hat{\psi}_{\bm{k}}^*} \right] + c.c. ,
\end{equation}
where $\hat{\psi}_{\bm k}$ is shorthand for $\hat{\psi} (\bm{k},t)$, and $c.c.$ denotes the complex conjugate. This equation results in the Grand-canonical distribution provided that $\beta F$ is a positive defined function. Thus, by solving numerically the Galerkin truncated Eq.~\eqref{eq: RGLET}, we can construct finite-temperature rotating equilibria.

When solving RGLET in many cases we will prefer to control the number of particles (or equivalently, the mean density $\Bar{\rho}$) instead of the chemical potential. This is equivalent to working in the Canonical ensemble. In practice, this can be done easily by solving Eq.~\eqref{eq: RGLET} coupled with an equation\cite{Krstulovic2011}
\begin{equation}
    \frac{\partial \mu}{\partial t} = - \alpha (\Bar{\rho} - \rho_0) ,
\end{equation}
where $\rho_0$ is the target (fixed) mean density in the center of the harmonic trap, and $\alpha$ controls how fast convergence to the desired mean density takes place.

\subsection{Condensate characteristic lengths}

Once a flow is created, several length scales can be used to characterize turbulence and its time evolution. A relevant characteristic scale in the study of classical turbulence is the flow integral scale $L_i$, which is associated to the large-scale flow correlation length. Let's write explicitly the isotropic incompressible kinetic energy spectrum, given by
\begin{equation}
    E_k^{\textrm{(i)}} (k) = \frac{1}{2} \int \Big| \widehat{[(\sqrt{\rho} \bm{v})^{\textrm{(i)}}]}_{\bm{k}} \Big|^2 \, k^2 d\Omega_{\bm{k}} ,
\end{equation}
where $\Omega_{\bm{k}}$ is the solid angle in Fourier space. We can then define the flow integral scale as
\begin{equation}
   \frac{L_i}{2 \pi} = \frac{\int E_k^{\textrm{(i)}} (k)/k \, dk}{\int E_k^{\textrm{(i)}} (k) \, dk} .
\end{equation}
From the Wiener–Khinchin theorem, it follows that this length is proportional to the flow correlation length obtained from the flow two-point spatial correlation function.

We can also define a spectrum of incompressible momentum, $P^\textrm{(i)}(k)$, given by\cite{Nore1997}
\begin{equation}
P^\textrm{(i)}(k) = \frac{1}{2} \int \Big| \widehat{[(\rho \bm{v})^{\textrm{(i)}}]}_{\bm{k}} \Big|^2 \, k^2 d\Omega_{\bm{k}} .
\end{equation}
It has been shown empirically that in a disordered tangle of vortices, for sufficiently large wave numbers $P^\textrm{(i)}(k)$ can be approximated as the momentum spectrum per vortex unit length of a single quantized vortex, $P_s^\textrm{(i)}(k)$, times the total length of the vortices\cite{Nore1997, Shukla2019}. Thus, the total vortex length $L_v$ can be estimated as
\begin{equation}
 \frac{L_v}{2\pi} = \frac{\int_{k_\textrm{min}}^{k_\textrm{max}} P^\textrm{(i)}(k)dk}{\int_{k_\textrm{min}}^{k_\textrm{max}} P_s^\textrm{(i)}(k)dk},
 \label{eq:vortex_length}
\end{equation}
where $k_\textrm{min}$ is a cutoff ($k_\textrm{min}=10$ in this study, as the contribution from smaller wave numbers is dominated by the trap geometry), and $k_\textrm{max}$ is the maximum resolved wave number in the simulations as before.

From $L_v$, the mean intervortex distance can be estimated as
\begin{equation}
 \ell = (\mathcal{V}/L_v)^{1/2}.
 \label{eq:intervortex}
\end{equation}

\subsection{Phase transitions}

Besides the phase transition from normal fluid to superfluid at $T_\lambda$ associated to the condensate, in recent years other phase transitions were discovered in classical and quantum fluids. It was found that the transition from a turbulent direct cascade of energy (i.e., from energy going from large to small scales) to an inverse energy cascade (i.e., from energy going from small to large scales, resulting in a flow self-organization and another condensate of the kinetic energy) can be a phase transition of second order, and controlled by the flow dimensionality. For condensates under GPE, it was also found that changing the aspect ratio of the condensate can trigger this transition \cite{Mller2020}. In classical rotating turbulence transitions were reported between direct cascades, flux loop states, and split cascades when varying the aspect ratio of the system or the Rossby number (i.e., the rotation speed) \cite{Clark_Di_Leoni_2020b}.

In the presence of rotation, a new threshold $\Omega_c$ appears for quantum fluids. Above this value, in order to minimize the free energy, the system can generate a vortex with one quantum of circulation along the axis of rotation. As rotation grows sufficiently strong, the flow turns quasi-2D, which could trigger the development of an inverse cascade in the turbulent regime. But as rotation grows, the system can also develop a vortex lattice. Thus, rotating quantum turbulence could suffer different transitions than classical rotating turbulence as $\Omega$ is varied. From the point of view of statistical mechanics, phase transitions can be discontinuous (first order), continuous but with discontinuous derivative (second order), or smooth with no symmetry breaking (sometimes called 'of infinite order'). It is yet unclear whether $\Omega_c$ or the flow two-dimensionalization resulting from rotation can result in phase transitions, and of what order.

\section{Numerical simulations}

\subsection{Numerical methods}

In the following we numerically solve Eq.~\eqref{eq:RARGLE} to generate initial conditions and then use Eq.~\eqref{eq:RGPE} to evolve the condensates in time.
Eq.~\eqref{eq: RGLET} is also numerically solved in order to generate thermalized states. In all cases we use an axisymmetric harmonic potential of the form $V(\bm{r}) = m \omega_\perp^2 (x^2+y^2)/2$ corresponding the limit of a vertically (infinitely) stretched cigar-shaped trap. This choice is done to reduce contamination of axisymmetric turbulent quantities (expected to be dominant in the rotating case) due to the trap geometry.

\begin{figure}
    \centering
    \includegraphics[width=8.5cm]{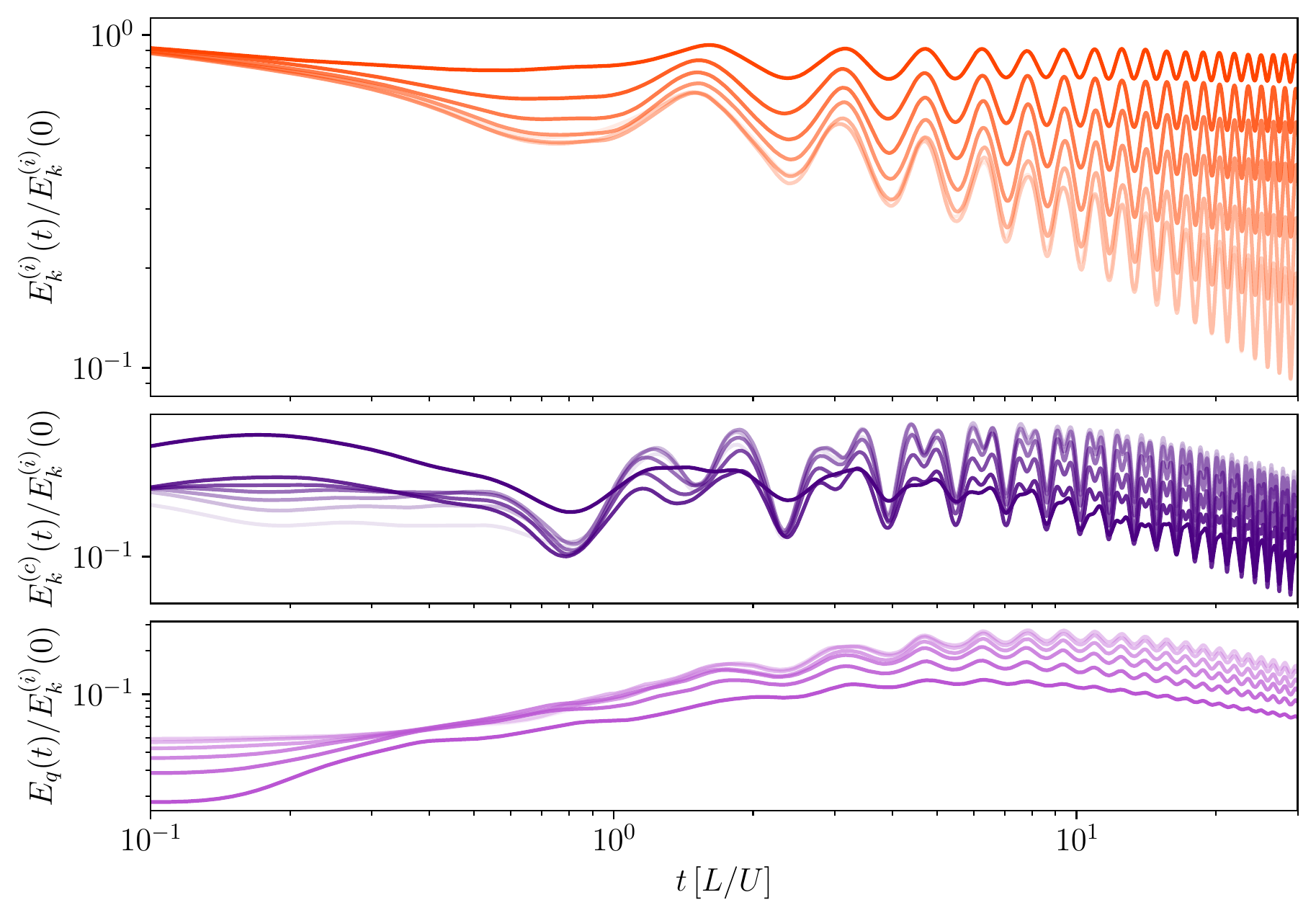}
    \caption{Time evolution of energy components for simulations with different values of $\Omega$ (from light, indicating $\Omega=0$, to darker colors, indicating increasing $\Omega$ up to $4.5 \Omega_c$. From top to bottom, incompressible kinetic energy, compressible kinetic energy, and quantum energy, all normalized by the incompressible kinetic energy at $t=0$.}
    \label{fig: energy evolution}
\end{figure}

To solve the equations we use a pseudospectral Fourier-based method in a spatial grid of $N^3 = 512^3$ points, with the $2/3$ rule for dealiasing, and a fourth-order Runge-Kutta method for time evolution for RGPE (an Euler time stepping method is used for the dissipative systems described by RARGLE and RGLET). In all cases we use the parallel code GHOST, which is publicly available \cite{Mininni2011}, in a cubic domain of dimensions $[-\pi,\pi]L \times [-\pi,\pi]L \times [-\pi,\pi]L$, so that the domain edges have length $2\pi L$. Periodic conditions are used in the $z$ direction. Because of the non-periodic nature of the angular momentum operator $J_z = -i \hbar(x \partial_y - y \partial_x)$ and of the potential $V(x,y)$ in the $x$ and $y$ directions, we smoothly extend them in a thin layer near the borders of the domain to make them (as well as all their spatial derivatives) periodic, allowing us to use Fourier expansions while controlling Gibbs phenomena\cite{Fontana_2020}. To do so, we perform a convolution between the Fourier transform of $V(\bm{r})$ or $J_z$ and a Gaussian filter in $k_x$ and $k_y$. This regularization is done in a region far away from the trap center, such that the density of the condensate in that region is negligible \cite{AmetteEstrada2022}. The width of the filter was chosen in an empirical way so that it minimizes the approximation error of $V({\bf r})$ and $J_z$ in the vicinity of the center of the trap. We used a width for the Gaussian filter $\sigma = (N \Delta k)/17 $, where $\Delta k$ is the resolution in wave number space. With this choice, errors in the computation of $V({\bf r})$ and $J_z$ were almost constant and $\approx 10^{-7}$ in the region occupied by the condensate. Values of $\omega_\perp$ were also chosen to keep the condensate confined in the region of the $xy$ plane satisfying these errors. With this method a standard pseudospectral method can be used to solve the equations, preserving the usual convergence properties of the method. The remaining parameters ($c$, $\xi$, $\rho_0$, and $\Omega$) were chosen as in the simulations reported in Ref.~ \cite{AmetteEstrada2022}. 

\begin{figure}
    \centering
    \includegraphics[width=8.5cm]{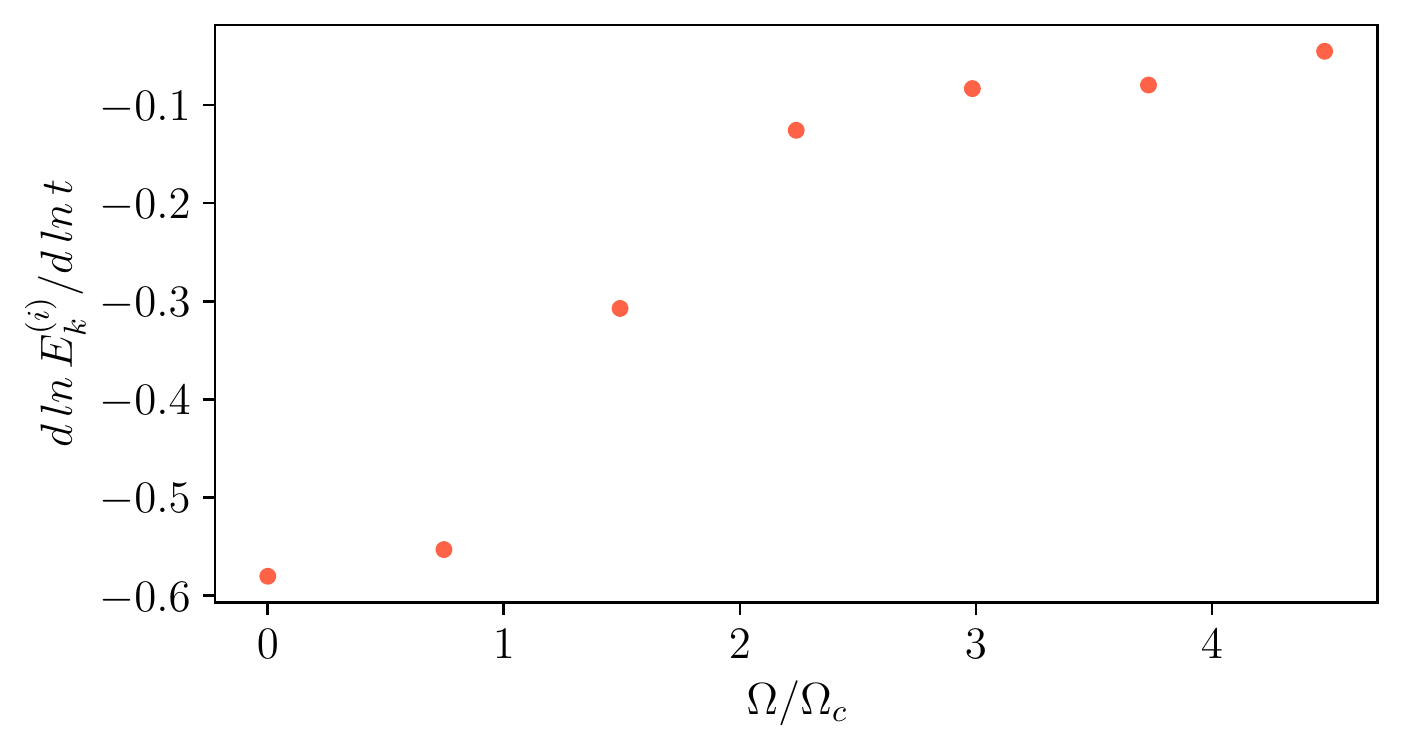}
    \caption{Exponent of the decay of the  incompressible kinetic energy with time, assuming a power law $E_k^{(i)}(t) \sim t^{\alpha}$, as a function of $\Omega/ \Omega_c$.}
    \label{fig: Einc exponents}
\end{figure}

\subsection{Preparation of initial conditions}

For the preparation of the initial conditions and finite temperature states, we must solve Eqs.~\eqref{eq:RARGLE} or Eq.~\eqref{eq: RGLET} respectively. In the latter case we start with a Gaussian density profile and let the system evolve until it reaches a steady state. In the former case we must choose a velocity field ${\bf u}$ to integrate Eq.~\ref{eq:RARGLE}. To that end we perturb a Gaussian density profile with a three dimensional random array of vortices using the initial conditions described in Ref.~\cite{Mller2020}. Parameters are chosen in such a way that the incompressible kinetic spectrum of the final state (i.e., of the initial conditions for the integration with RGPE) has a peak at $k \approx 5$. In this way, room is left in spectral space for the system to self-organize at large scales.

\section{Rotating Turbulence and thermalization}

\begin{figure}
    \centering
    \includegraphics[width=8.5cm]{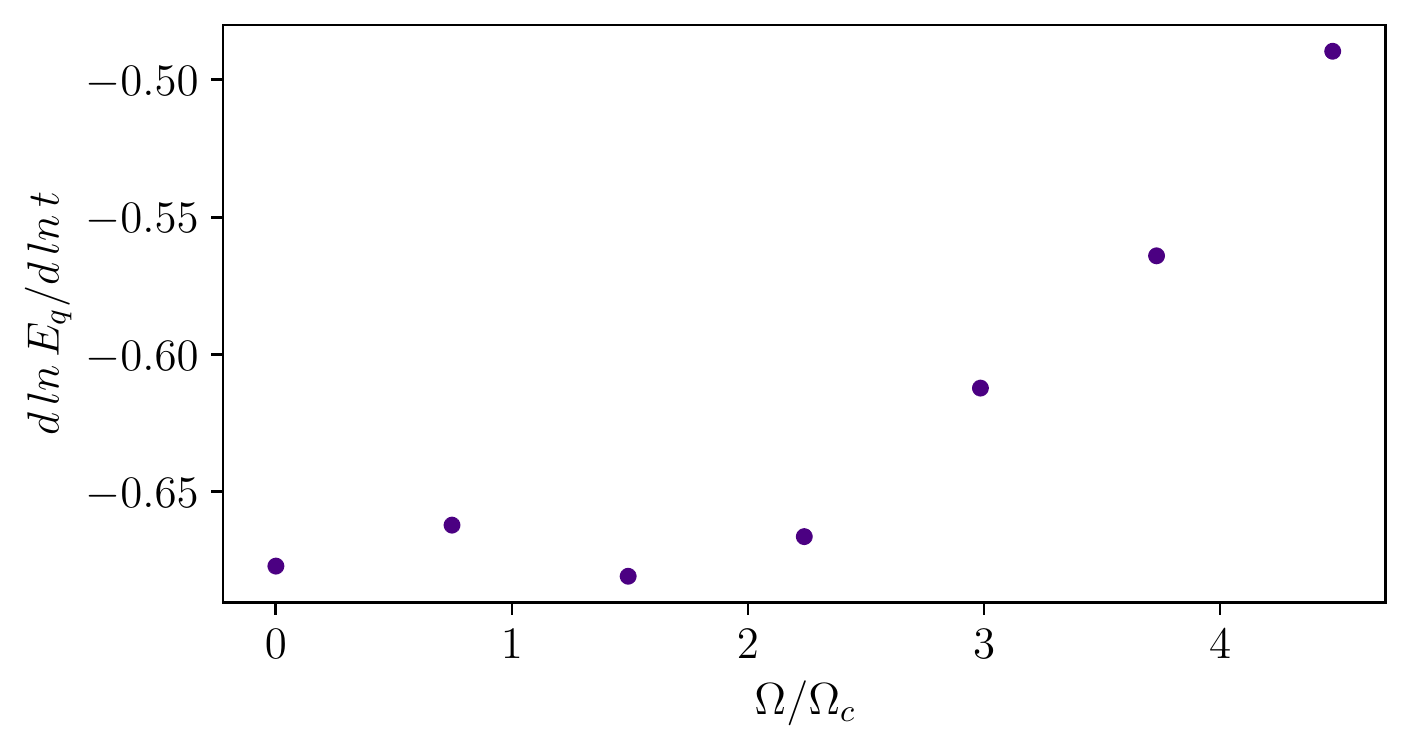}
    \caption{Decay exponents of the quantum energy at late times in all simulations, assuming a power law $E_q(t) \sim t^\beta$, as a function of $\Omega / \Omega_c$}
    \label{fig: quantum exponents}
\end{figure}

\subsection{Freely decaying flows}

Global quantities provide useful insights into the evolution of turbulent systems. In this section we begin by studying the evolution of energy components and flow characteristic scales as turbulence decays freely, to undersand how the evolution of quantum turbulence is affected by rotation. To that end we integrate for long times RGPE with the turbulent initial conditions just described, for different values of $\Omega$. Parameters for the simulations are the same as in Ref.~\cite{AmetteEstrada2022}.

We first analyze the general energy budget in the system. Equation \eqref{eq:RGPE} is conservative, so that changes in the total energy are only due to the numerical errors, and negligible in our case. Thus, when several components of the energy decrease in time, others must grow to keep the total energy constant. As a rule, the incompressible kinetic energy decreases in all runs as turbulence decays, and either the compressible kinetic energy (i.e., sound), internal, or rotational energy must increase as the turbulent kinetic energy is dissipated and transformed into disorder or rigid body rotation. Figure \ref{fig: energy evolution}  shows the evolution of the incompressible, compressible, and quantum energy components as a function of time, for several simulations with $\Omega$ between $0$ and $1.2 \, U/L$ (where $U$ is a unit velocity). Before discussing the effect of rotation on these quantities, note that all energy components display oscillations with a frequency independent of $\Omega$, which correspond to the breathing mode of the condensate in the trap. The frequency is proportional to $2 \omega_\perp$, in agreement with theoretical predictions \cite{Stringari1996}. 

When $\Omega = 0$ in Fig.~\ref{fig: energy evolution} (lightest curves), the envelope of the incompressible kinetic energy (i.e., ignoring the breathing mode oscillations)  decreases in time in a fashion compatible with a power-law decay, while the compressible and quantum energies first grow and then decay. Eventually the internal energy grows to compensate for these decays (not shown). This is compatible with the classical behavior of freely decaying turbulence, where kinetic energy cascades to smaller scales until it is eventually converted into internal energy. In this case, the early growth of compressible and quantum energies corresponds to the emission of sound waves and to the development of density inhomogenities. However, as $\Omega$ increases (darker curves) the incompressible kinetic energy decays at a slower rate until it remains almost constant for large enough $\Omega$ (except for the breathing mode oscillations). As a result, very little increase of the compressible and quantum energies is observed at early times (at later times, both the internal and rotational energies increase to compensate for the small decay of the other energy components). This latter behavior indicates a substantial decrease in the energy dissipation by turbulence as rotation increases.

\begin{figure}
    \centering
    \includegraphics[width=8.5cm]{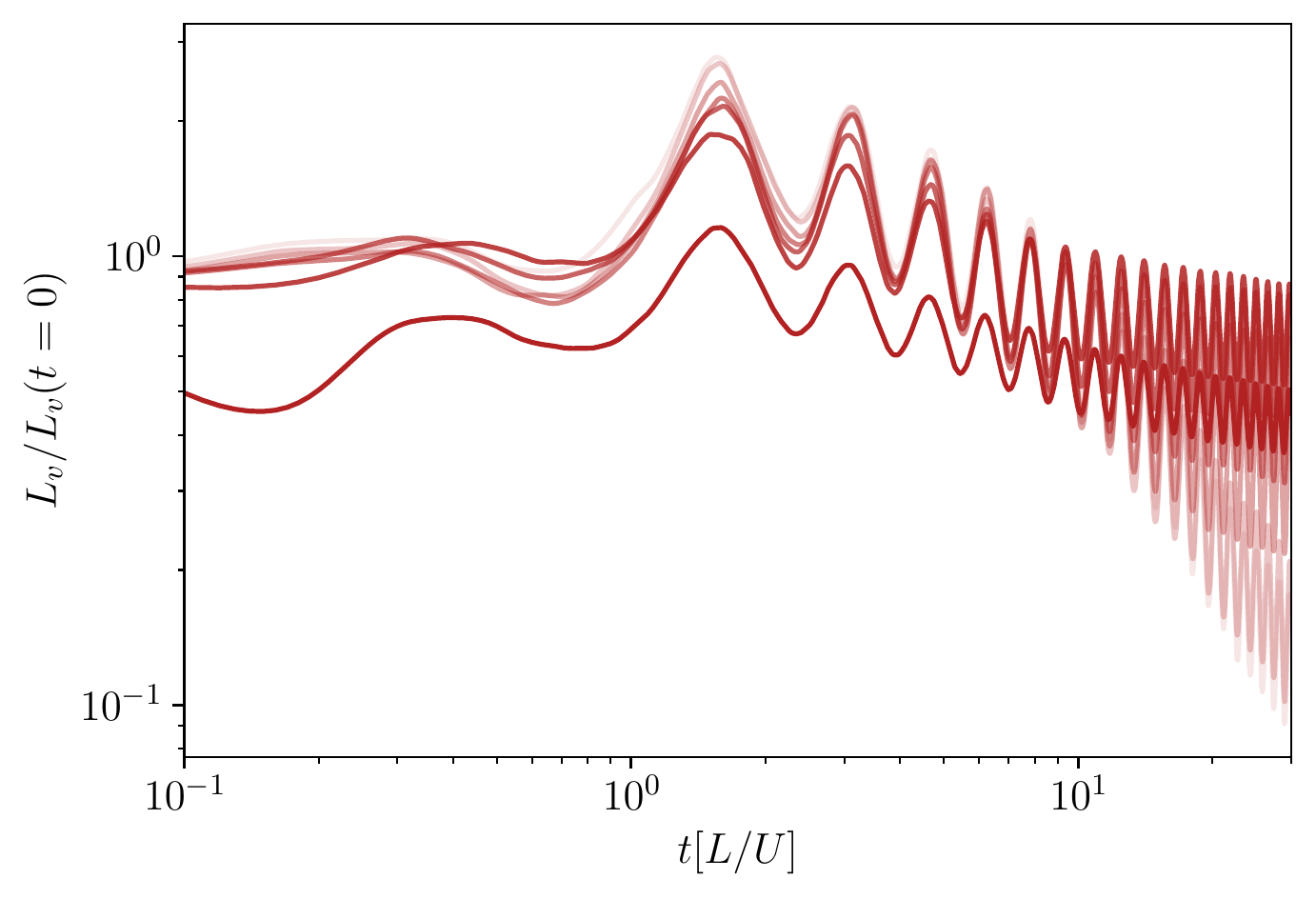}
    \includegraphics[width=8.5cm]{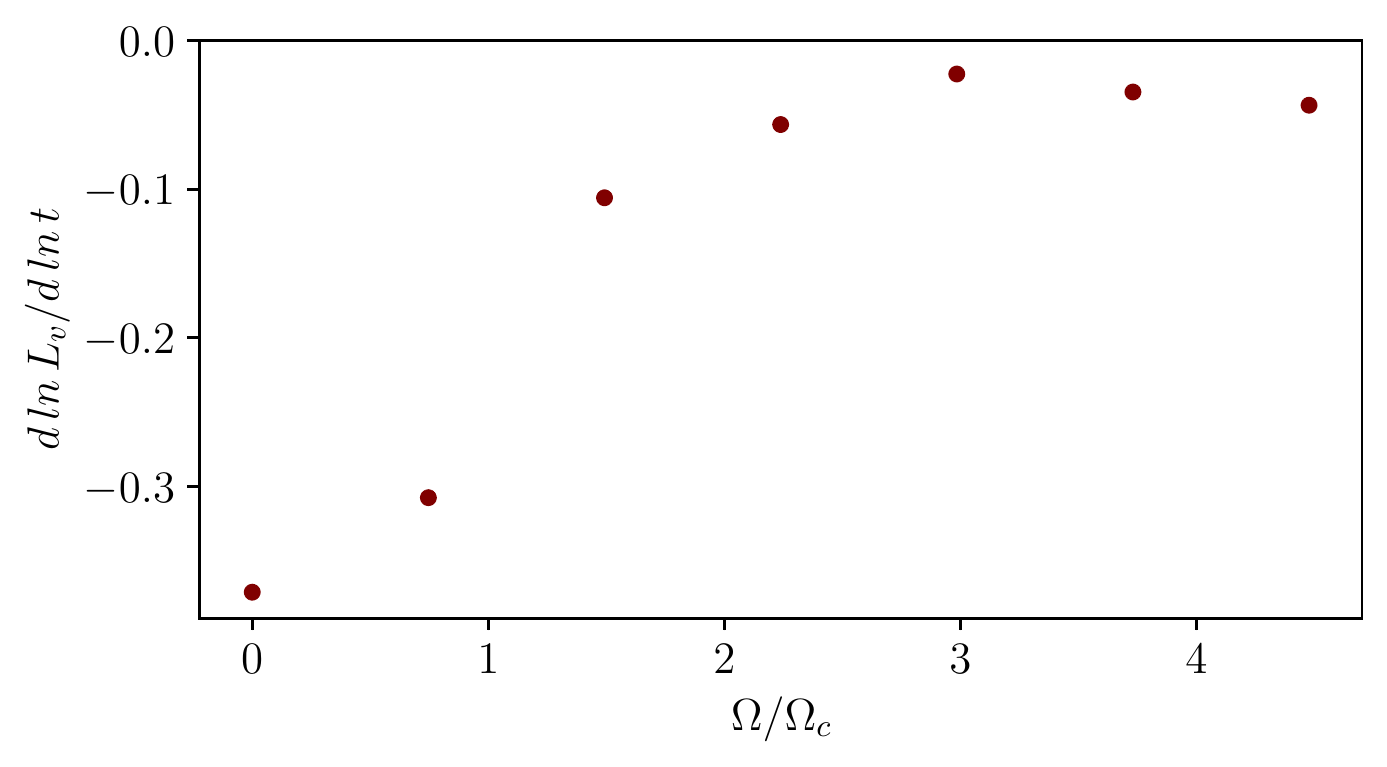}
    \caption{{\it Top:} Evolution of the total vortex length as a function of time, normalized by its value at $t=0$, as a function of time. From light to dark, $\Omega=0$ up to $4.5 \Omega_c$. {\it Bottom:} Decay exponents of the vortex length at late times, as a function of $\Omega / \Omega_c$.}
    \label{fig: vortex length}
\end{figure}

A more quantitative description of this process can be obtained by assuming, as often done in turbulent flows, that energies evolve in time in a self-similar way as $\sim t^\alpha$, for times $t > t_0$ where $t_0$ marks the time of the beginning of the self-similar decay. The exponent $\alpha$ ($<0$ for the decaying case) can then be computed from the logarithmic derivative, i.e., $d \ln E_k^{(i)}(t)/d \ln t$. To avoid contamination from the breathing mode in Fig.~\ref{fig: energy evolution}, we take only the amplitude of each peak for all times $t>t_0$. The decay exponent for the incompressible kinetic energy as a function of the dimensionless control parameter $\Omega / \Omega_c$ is shown in Fig.~\ref{fig: Einc exponents}. Note the exponent remains approximately constant for $\Omega / \Omega_c <1$, and grows rapidly for $\Omega / \Omega_c >1$ reaching saturation in a small value close to zero (i.e., for almost non-decaying incompressible kinetic energy). Thus, there are two regimes separated by $\Omega_c$. Below this threshold rotation is too weak to affect the system (as its associated circulation is smaller than the circulation needed to create one quantized vortex). Turbulence evolves as in the non-rotating case, with a Kelvin wave cascade of energy towards smaller scales that results in incompressible kinetic energy dissipation. Sufficiently above $\Omega_c$ the incompressible kinetic energy remains almost constant, and the direct cascade seems to be quenched. As we will see later, this is associated with the development of a self-organization process in which energy is transferred instead towards larger scales.

The same analysis can be done with the quantum energy at late times (no clear power law scaling was found for the compressible kinetic energy, which grows first, albeit less for larger $\Omega$, and then slowly decays). The decay exponent of $E_q(t)$, i.e., the logarithmic derivative of the amplitude of its peaks $d \ln E_q(t)/d \ln t$, as a function of $\Omega/\Omega_c$, is shown in Fig.~\ref{fig: quantum exponents}. The exponent is always negative but, as in the case of the incompressible kinetic energy, it shows a distinct behaviour as $\Omega / \Omega_c$ increases. As rotation grows above $\Omega_c$ the exponent decreases (in absolute sign), this time monotonically with increasing $\Omega$. This is consistent with a slower decay and, together with the results shown in Fig.~\ref{fig: energy evolution}, with the generation of weaker density inhomogeneities as $\Omega$ increases.

\begin{figure}
    \centering
    \includegraphics[width=8.5cm]{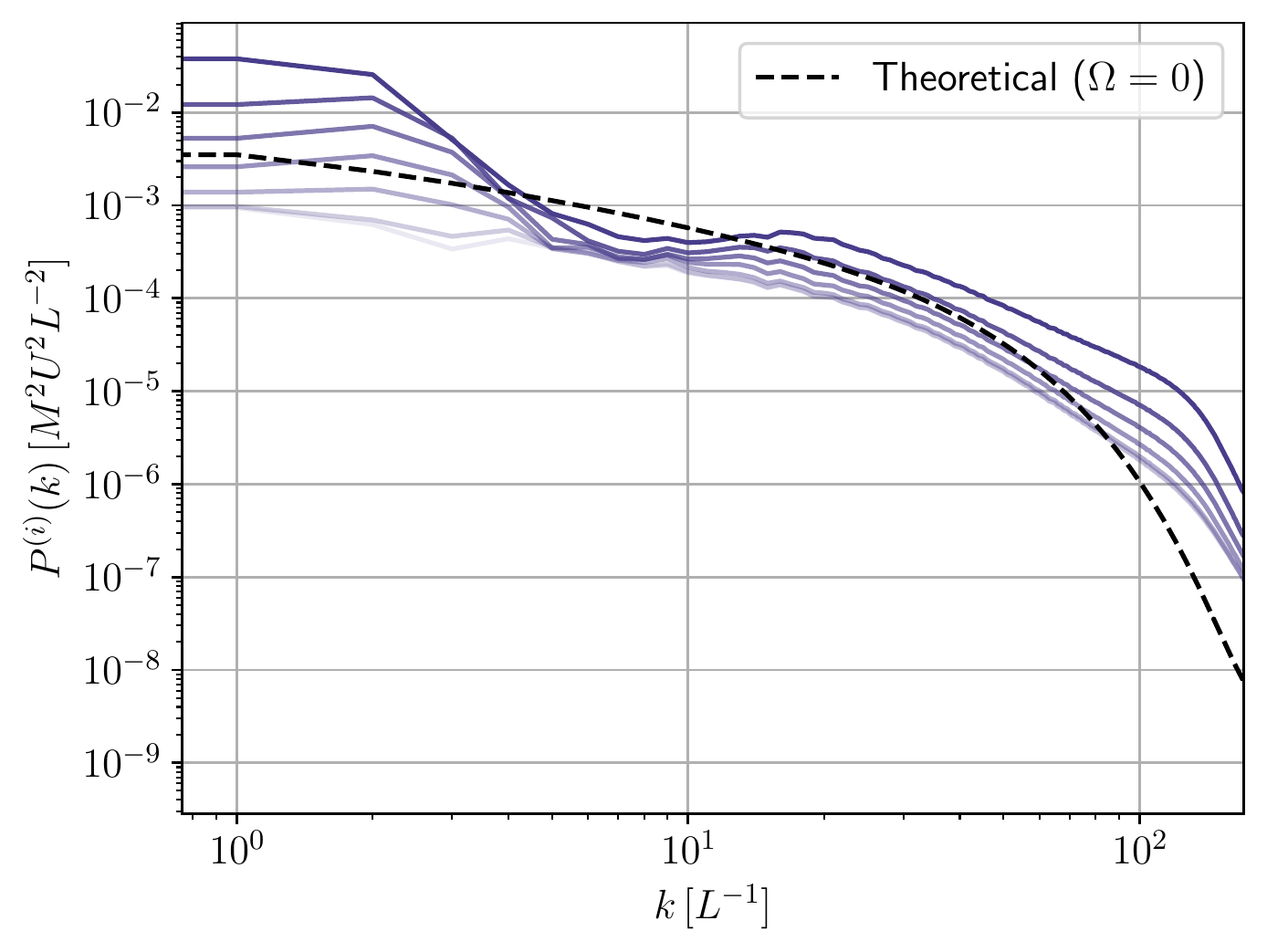}
    \caption{Momentum spectrum at late times for several simulations with different values of $\Omega$ (with darker colors indicating larger $\Omega$). The dashed black line indicates the theoretical momentum spectrum per unit length of one vortex \cite{Nore1997} for $\Omega=0$, multiplied by the total vortex length in the simulation with $\Omega = 0$.}
    \label{fig: momspectrums}
\end{figure}

The time evolution of characteristic length scales give further insights on the system dynamics. The total vortex length as a function of time for different simulations is shown in the top panel of Fig.~\ref{fig: vortex length}. As turbulence evolves, vortex stretching results in a growth of the total vortex length at early times, followed by decay compatible with a power law as the flow decays. However, rotation strongly affects this evolution. For sufficiently large $\Omega$ (darker curves) $L_v$ grows less at early times (i.e., there is less vortex stretching, compatible with the behavior expected in a quasi-2D flow), and its decay is slower at late times. This is further confirmed by the decay exponent computed from the amplitude of all the peaks of $L_v$ at late times, with is shown as a function of $\Omega / \Omega_c$ in the bottom panel of Fig.~\ref{fig: vortex length}. The exponent again changes around $\Omega / \Omega_c \approx 1$. For $\Omega / \Omega_c > 1$ the exponent approaches zero, confirming that the total vortex length (and thus also the intervortex separation) change less for sufficiently large rotation rates.

\begin{figure}
    \centering
    \includegraphics[width=8.5cm]{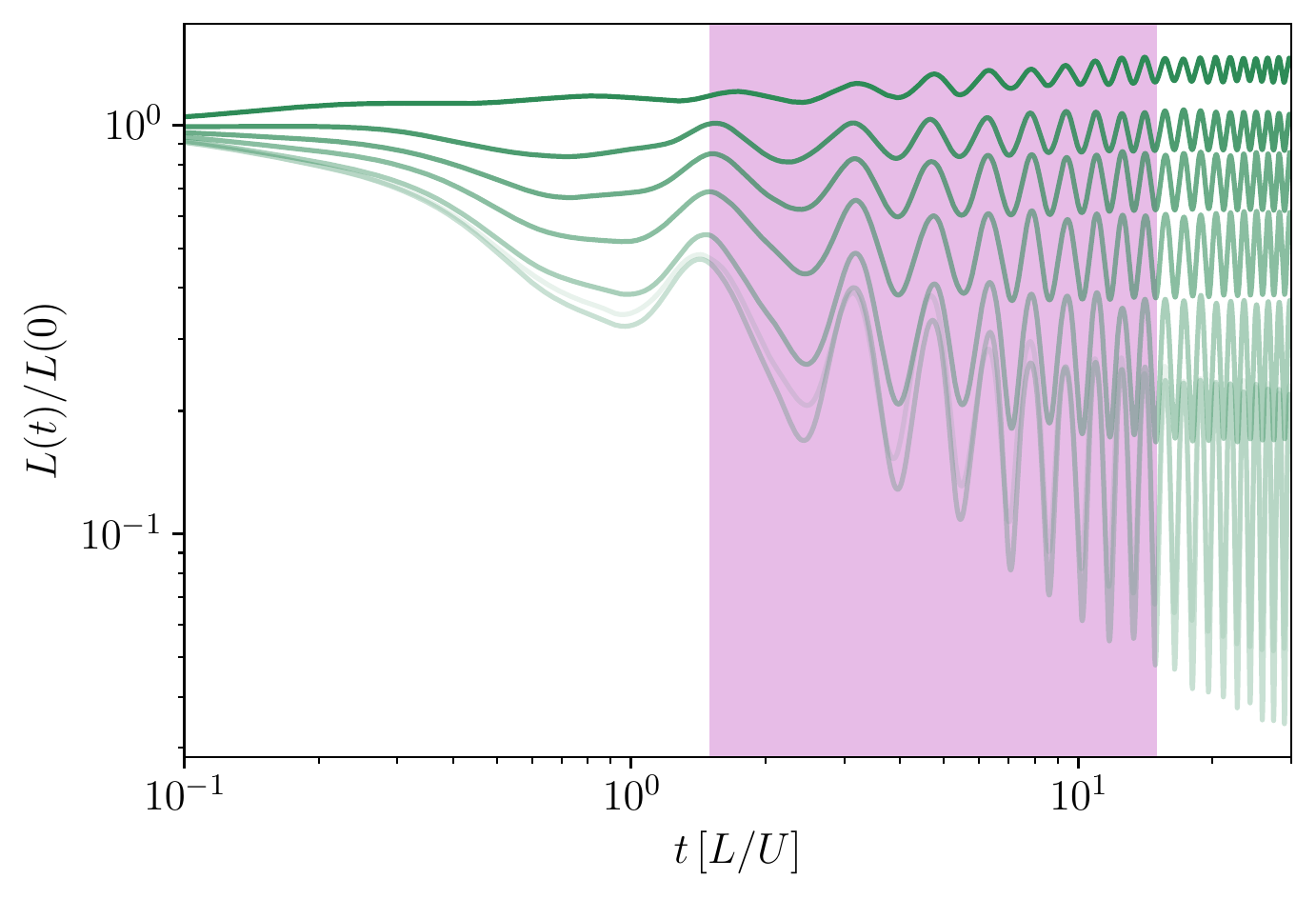}
    \includegraphics[width=8.5cm]{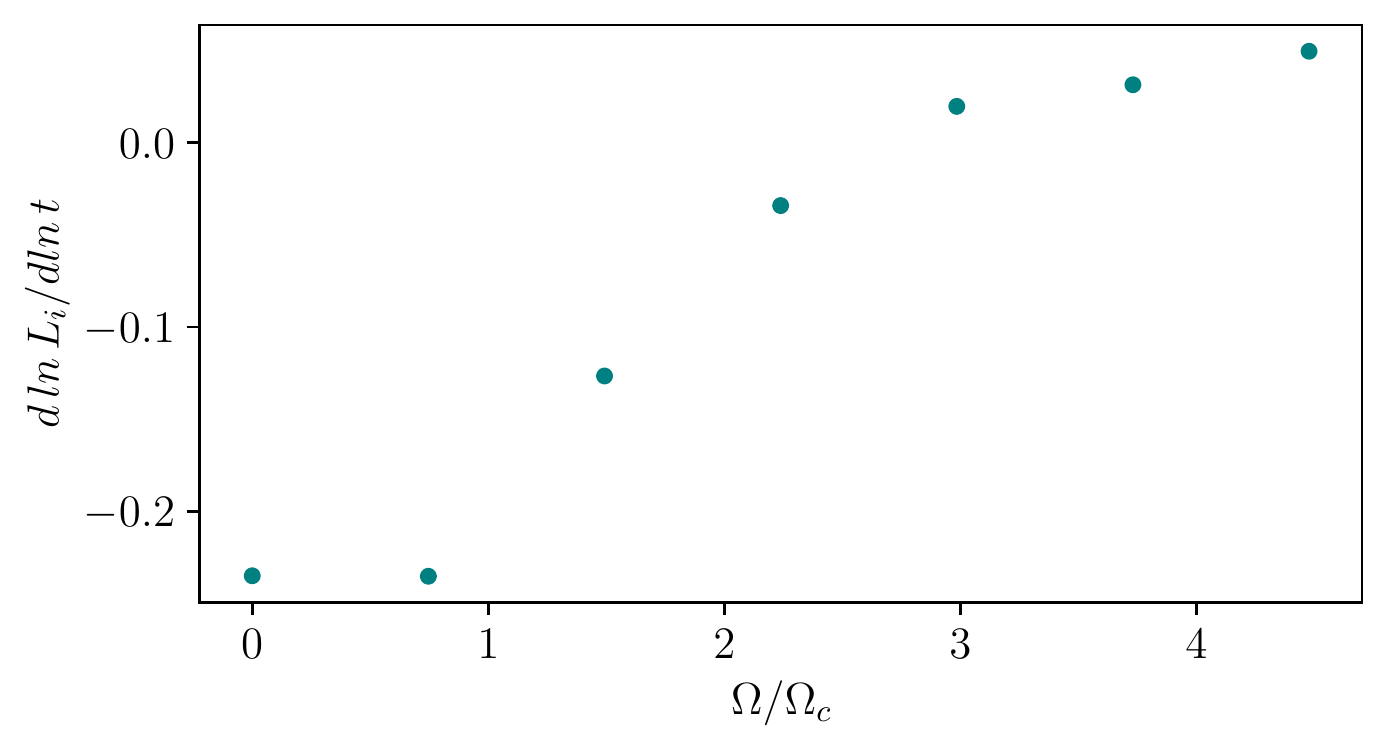}
    \caption{{\it Top:} Time evolution of the flow integral length scale $L_i(t)$ (normalized by its initial value at $t=0$) for simulations with different values of $\Omega$ (from light to dark curves, $\Omega=0$ to $4.5 \Omega_c$). The shaded region indicates the times for which a power-law was adjusted. {\it Bottom:} Decay exponent of the integral length scale, assuming $L_i(t) \sim t^\gamma$, as a function of $\Omega / \Omega_c$.}
    \label{fig: integral length}
\end{figure}

How is the spectral distribution of these vortices? A first hint is given by the incompressible momentum spectrum at late times in all simulations, shown in Fig.~\ref{fig: momspectrums}. Remember that the total vortex length is the integral of this spectrum properly normalized, as in Eq.~\eqref{eq:vortex_length}, and that the mean intervortex distance also follows from this spectrum, as defined in Eq.~\eqref{eq:intervortex}. Figure \ref{fig: momspectrums} also shows as a reference the theoretical incompressible momentum spectrum per unit length of one vortex \cite{Nore1997} for $\Omega=0$, multiplied by the total vortex length in the simulation with $\Omega=0$. Thus, differences from this spectrum (and differences with the spectrum of the numerical simulation with $\Omega=0$) indicate differences from an isotropic and disordered vortex tangle when rotation is present. The spectrum continually changes as $\Omega$ is increased. The two most conspicuous differences can be found at large and at small wave numbers: On the one hand, for large $\Omega$ the spectrum has more power at large values of $k$. On the other hand, a clear accumulation of momentum at low wave numbers (i.e., at large scales) can be seen. A gap in the spectrum separates these two features. Thus, the quantized vortices for $\Omega>\Omega_c$ seem to have a large-scale ordering with small-scale fluctuations that are different from the disordered vortex bundle when $\Omega<\Omega_c$.

\subsection{Early time inverse energy transfer}

\begin{figure}
    \centering
    \includegraphics[width=8.7cm]{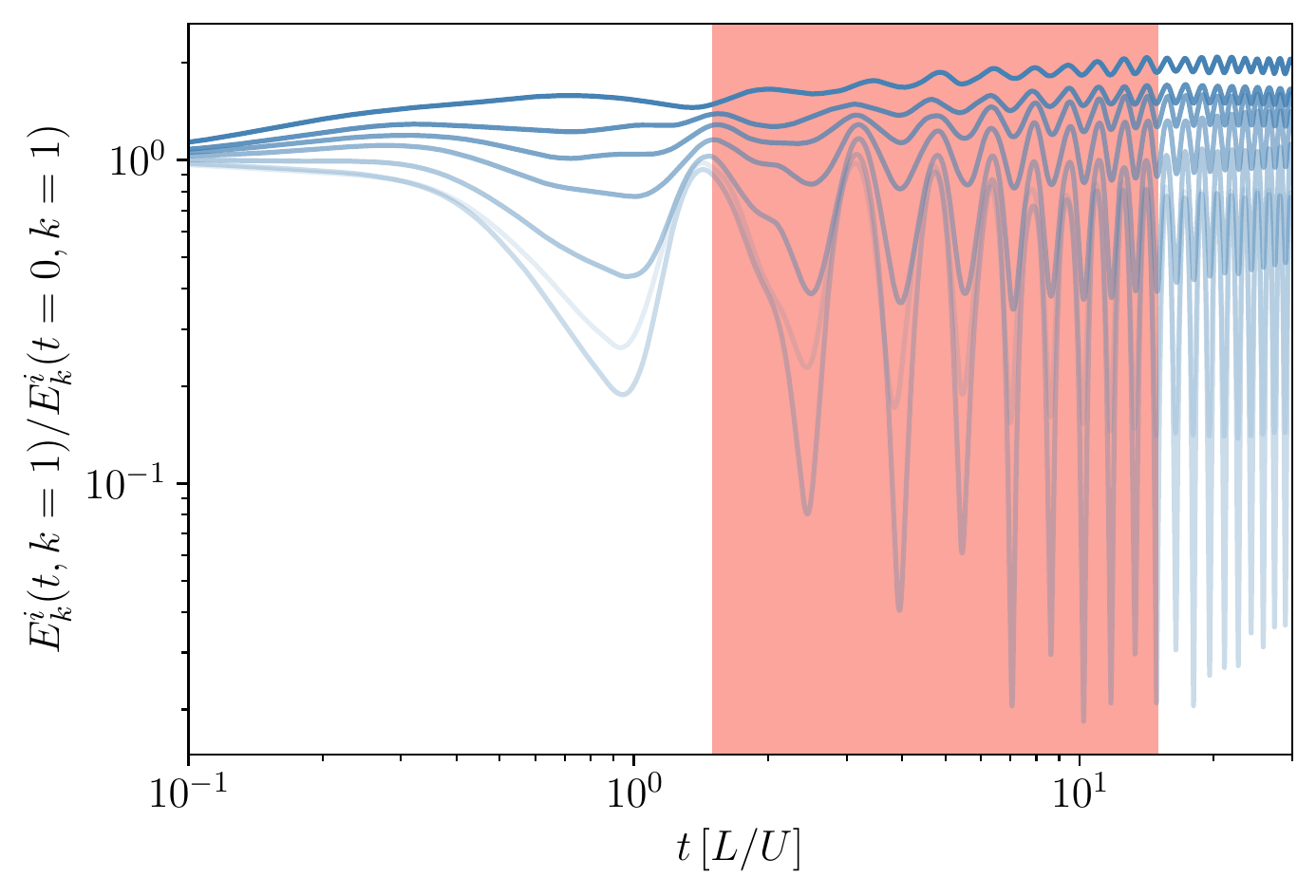}
    \includegraphics[width=8.7cm]{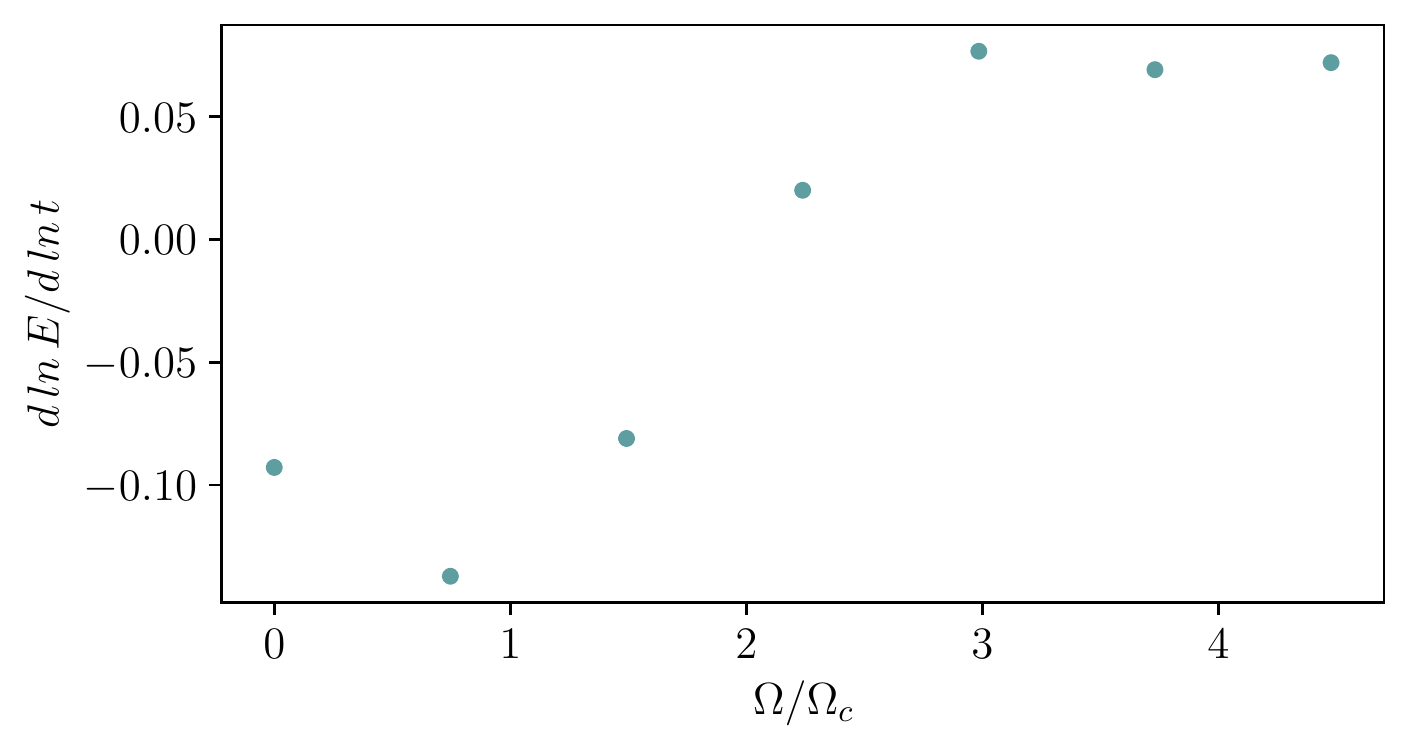}
    \caption{{\it Top:} Time evolution of the incompressible kinetic energy in the $k=1$ Fourier shell, normalized by its initial value at $t=0$, for simulations with different values of $\Omega$ (from light to dark curves, $\Omega=0$ to $4.5 \Omega_c$). The shaded region indicates the times for which a power-law was adjusted. {\it Bottom:} Decay exponent as a function of $\Omega / \Omega_c$. Note $-dE_k^{(i)}(k=1)/dt$ is the energy flux at that wave number.}
    \label{fig: energy k = 1}
\end{figure}

The early time behavior of $L_v(t)$ (i.e., the significant decrease of vortex stretching in the rotating BECs), together with the very slow decay of the incompressible kinetic energy and the accumulation of momentum at small wave numbers for large values of $\Omega$ in Figs.~\ref{fig: energy evolution} and \ref{fig: momspectrums}, suggest that the incompressible kinetic energy may be accumulating at large scales in the condensate, or in other words, that the incompressible kinetic energy could be cascading inversely to larger scales. The study of the time evolution of large-scale quantities in the flow at earlier times (between $t=1.5$ and $15$) confirms that this is indeed the case and provides information on this process. The choice of analyzing the growth or decay of large-scale quantities at intermediate times is justified by the fact that the turbulence is freely decaying, and a self-organization process can only be sustained for as long as turbulence remains strong enough, together with the fact that once the energy has reached the largest available scale in the condensate, the process is expected to saturate in systems with finite size. This behavior will be explicitly confirmed below.

The evolution of the flow integral length scale as a function of time and its logarithmic derivative are shown in Fig.~\ref{fig: integral length}. In the top panel, the time evolution shows a different behavior as $\Omega$ increases. While in simulations with small $\Omega$ the envelope of $L_i(t)$ decreases in time (as expected in freely decaying classical three dimensional turbulence \cite{Teitelbaum2010}), for $\Omega<\Omega_c$ the envelope of $L_i(t)$ first increases and then saturates at a constant value. The bottom panel in Fig.~\ref{fig: integral length} shows the power-law exponent of the growth or decay of $L_i(t)$, assuming $L_i (t) \sim t^\gamma$. It is approximately constant and negative for $\Omega<\Omega_c$, decreases rapidly for $\Omega>\Omega_c$, and becomes positive (i.e., $L_i$ grows with time) for $\Omega \gtrsim 3 \Omega_c$. Thus, the flow integral length scale transitions from decaying to growing as the rotation rate increases. Note that $L_i(t)$ is a measure of the large-scale flow correlation, and thus its growth indicates the system creates larger vortex structures than those present in the initial conditions.

\begin{figure}
    \centering
    \includegraphics[width=8.5cm]{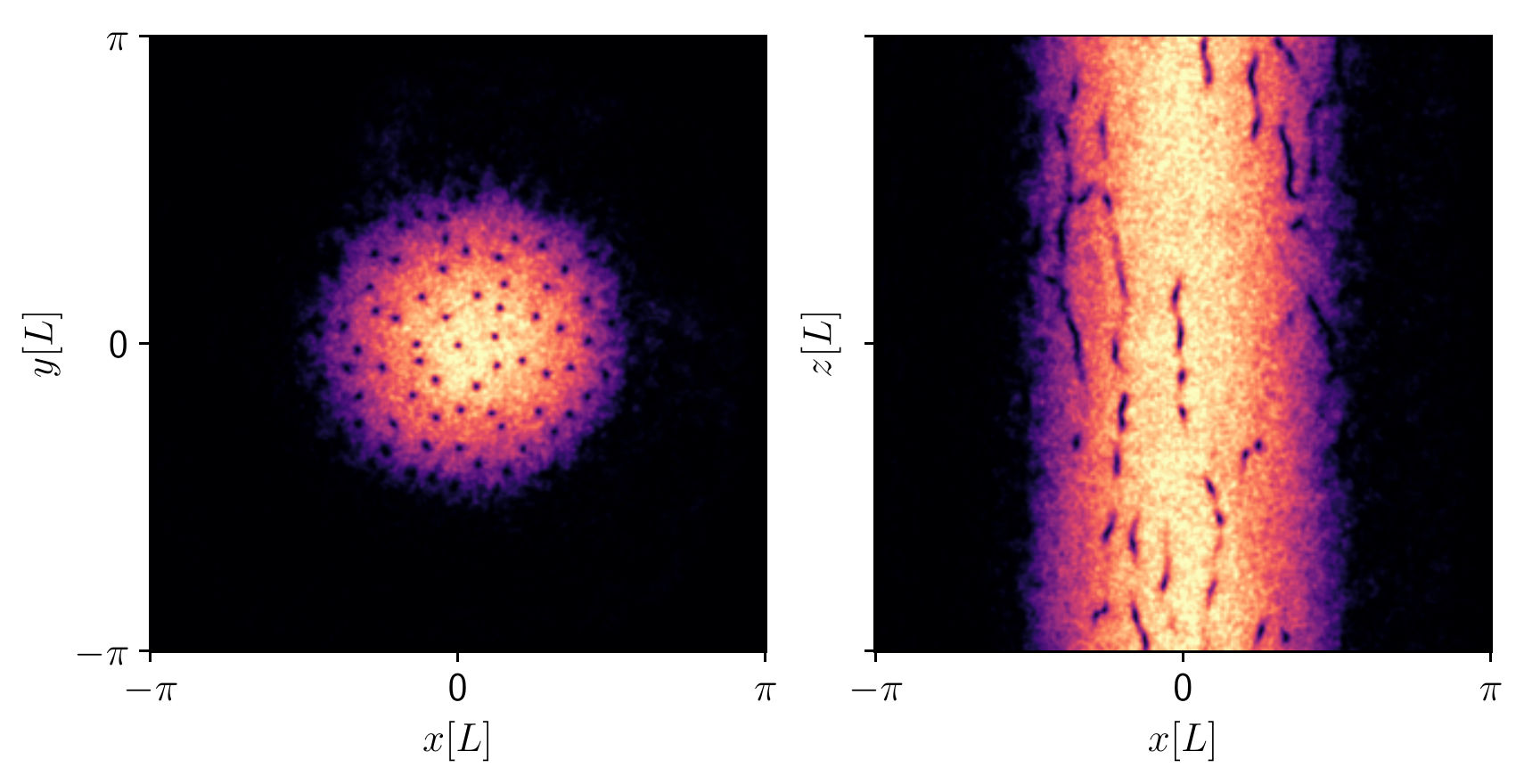}
    \caption{Two-dimensional slices of the mass density in a plane perpendicular to the rotation axis, $\rho(x,y,z=0)$ {\it (left)}, and in a plane parallel to the same axis, $\rho(x,y=0,z)$ {\it (right)}, at late time in a RGPE simulation with $\Omega = 1.2 \, U/L = 4.5 \Omega_c$.}
    \label{fig: realdensity_turbulence_decay}
\end{figure}

Figure \ref{fig: energy k = 1} shows the same quantities for the incompressible kinetic energy in the gravest modes (i.e., those in the $k=1$ Fourier shell), for simulations with different values of $\Omega$. We know from Fig.~\ref{fig: energy evolution} that for sufficiently large $\Omega$ the incompressible kinetic energy remains almost constant. The top panel in Fig.~\ref{fig: energy k = 1} shows that the way the system prevents this decay is by accumulating kinetic energy at the largest scales. For sufficiently large $\Omega$, the incompressible kinetic energy condensates in the gravest mode, resulting in the aforementioned growth of the flow integral scale. This is the result of an inverse transfer of energy, which is a feature of rotating quantum turbulence \cite{AmetteEstrada2022}. The bottom panel shows the logarithmic derivative of the envelope of this quantity, as a function of $\Omega/\Omega_c$. Again a transition is observed for $\Omega/\Omega_c \approx 1$, and $d \ln E_k^{(i)}(k=1,t)/d \ln t$ becomes positive when $\Omega$ is above $2 /\Omega_c$. Note also that $-dE_k^{(i)}(k=1)/dt$ is the energy flux at that wave number, and a positive value of the logarithmic derivative confirms energy is flowing in spectral space towards larger scales. Thus, as rotation increases, incompressible kinetic energy accumulates in small wave numbers, its transfer towards smaller scales is quenched as in classical rotating turbulence \cite{Waleffe1992, Waleffe_1993}, and the associated emission of compressible excitations diminishes.

\begin{figure}
    \centering
    \includegraphics[width=8.5cm]{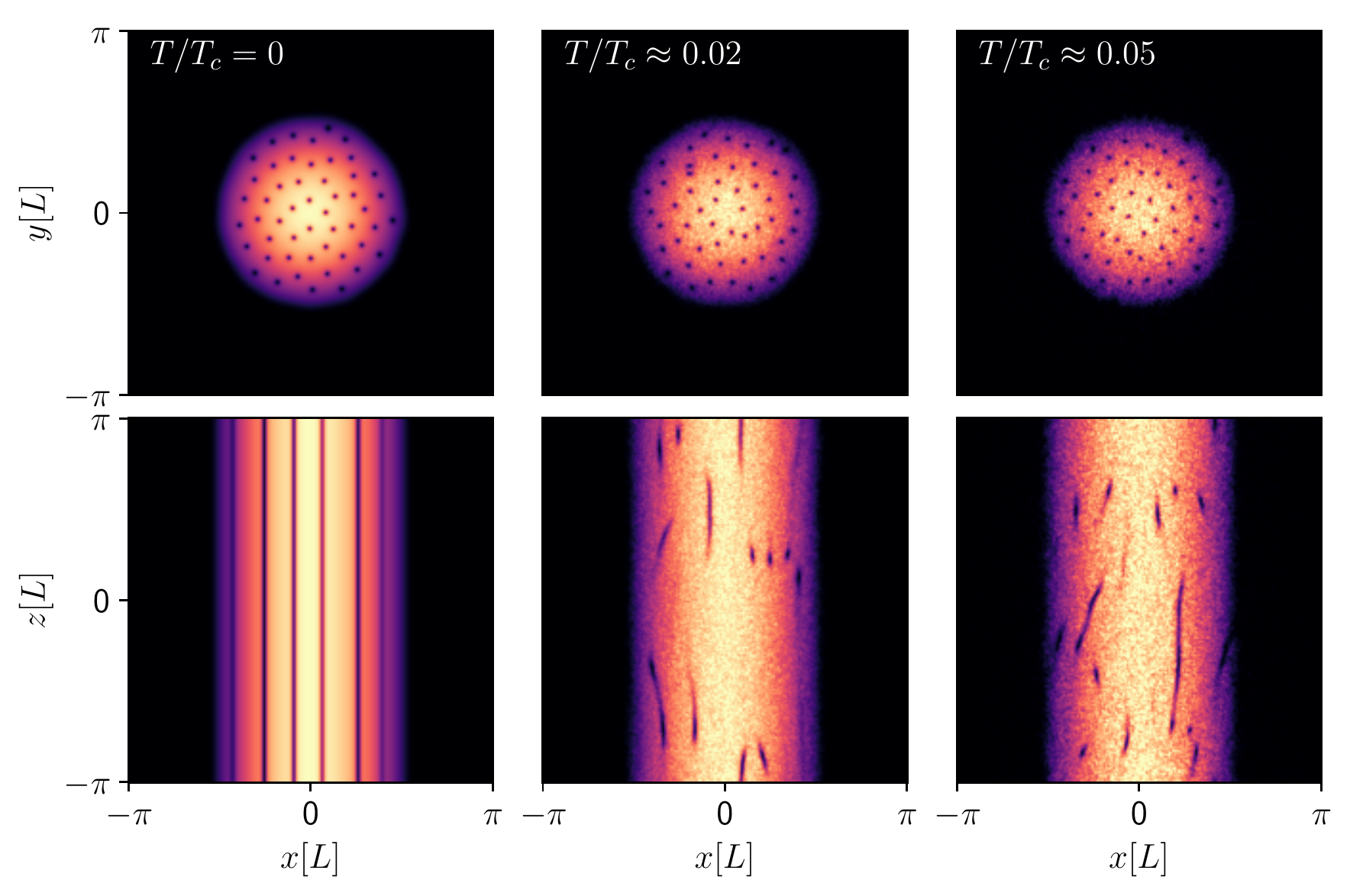}
    \caption{Two-dimensional slices of the mass density in a plane perpendicular to the rotation axis, $\rho(x,y,z=0)$ {\it (top)} and parallel to that axis, $\rho(x,y=0,z)$ {\it (bottom)} for thermalized states with increasing temperatures, in simulations with $\Omega = 1.2 \, U/L$.}
    \label{fig: realdensity_temperature}
\end{figure}

\subsection{Thermalized Abrikosov states}

What is the final state to which rotating turbulence decays? We have seen that for $\Omega > \Omega_c$ the incompressible kinetic energy decays more slowly (remaining almost constant for sufficiently large $\Omega$), the compressible kinetic energy grows less, vortex stretching decreases substantially, and a large-scale flow develops. Thus, even though we intuitively expect turbulence to generate a disordered state, in the presence of rotation some order may survive even for very long times. And the resulting state can be expected to be very different from the final stages of freely decaying classical rotating turbulence \cite{Teitelbaum2010}.

As time evolves and the flow decays to a new equilibrium, it is natural for structures that do not satisfy the system symmetries to vanish. From classical turbulence we know that rotation makes the flow quasi-2D \cite{Waleffe1992, Waleffe_1993, Cambon_1997, Cambon_2004}, with vortical structures that are predominantly aligned with the rotation axis. Figure \ref{fig: realdensity_turbulence_decay} shows 2D slices of the mass density at late time in RGPE simulations of freely decaying turbulence with $\Omega = 1.2 \, U/L = 4.5 \Omega_c$. Note that quantized vortices are predominantly aligned with the rotation axis, and axisymmetry is somehow recovered in a statistical sense. However, vortices are not perfectly parallel. Seen from top (i.e., in the $x$-$y$ plane) the vortices resemble an Abrikosov lattice but with random fluctuations. Seen from the side (e.g., in the $x$-$z$ plane), vortices wiggle and are not perfectly parallel. Waiting for longer times does not result in a cleaner or more ordered lattice. The reason is associated with the time evolution discussed previously. As time evolves, despite the transfer of incompressible kinetic energy towards large scales, some of the incompressible kinetic energy is transferred to smaller scales where it can excite phonons that partially thermalize \cite{AmetteEstrada2022}.

We can compare this final state with thermalized equilibria generated by RGLET in Eq.~\eqref{eq: RGLET}. In Fig.~\ref{fig: realdensity_temperature} the mass density of simulations of RGLET with different temperatures is shown. Temperature is measured in dimensionless units of $T/T_c$, where the critical temperature $T_c$ to get a condensate is obtained by scanning the solutions of RGLET for different temperatures using the method described in \cite{Shukla2019}. For $T=0$ a clear Abrikosov lattice is obtained. As temperature is increased, larger fluctuations appear and vortices are not perfectly aligned with the rotation axis. Note that these solutions are not obtained as a result of the decay of an initial flow. The system is started with a Gaussian density profile with no vortices, and the Abrikosov lattice appears as a solution that minimizes the system free energy at a given temperature: these are thermal equilibria superimposed with the Abrikosov lattice needed by the BEC to mimic the solid body rotation.

\begin{figure}
    \centering
    \includegraphics[width=8.5cm]{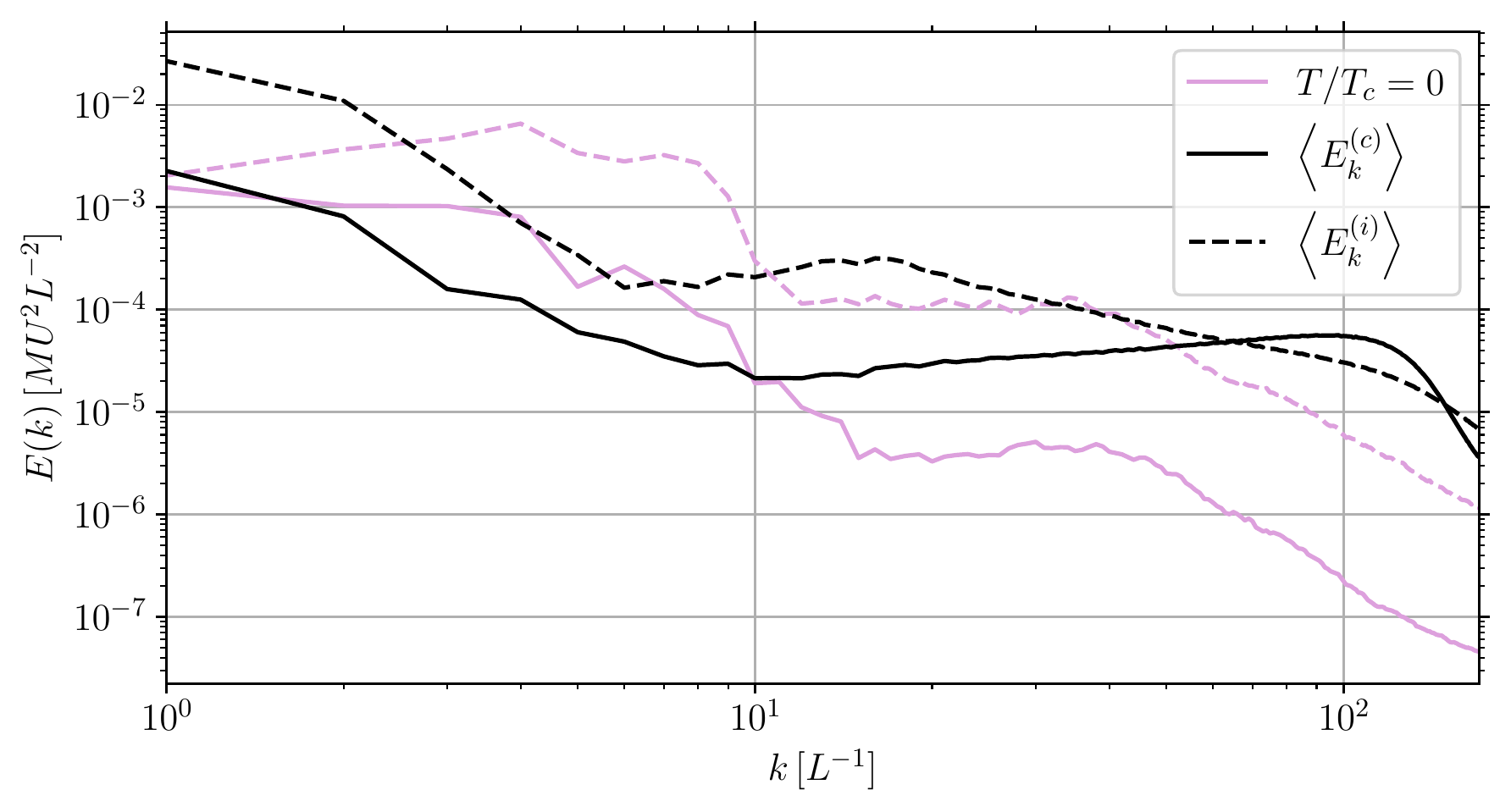}
    \includegraphics[width=8.5cm]{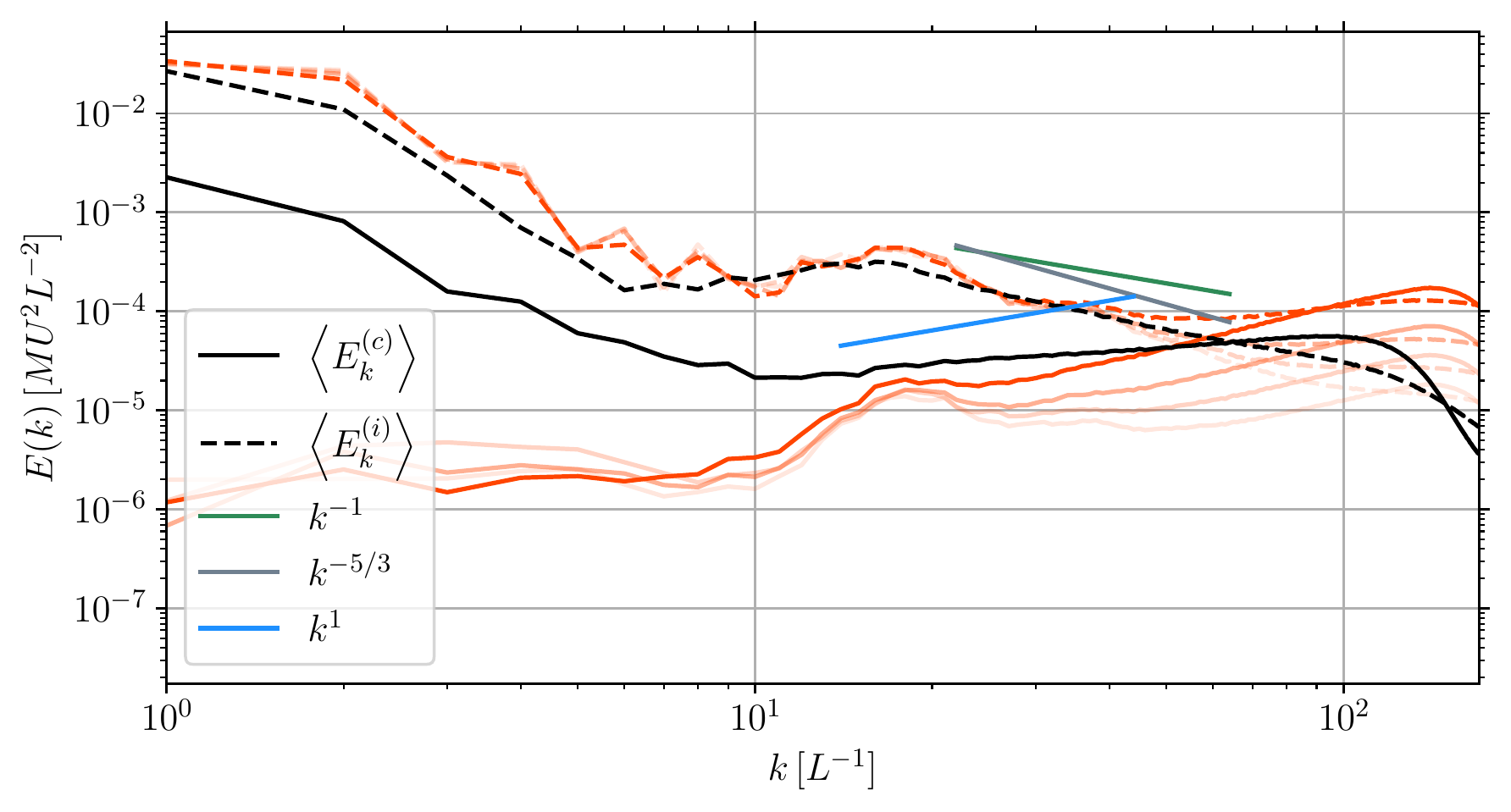}
    \caption{Isotropic compressible (solid) and incompressible (dashed) energy spectra for $\Omega = 1.2 \, U/L$, of: \textit{(Top)} The Abrikosov lattice at $T = 0$, compared with a RGPE late time solution, and \textit{(bottom)} different Abrikosov lattices at different $T > 0$, compared against the same RGPE solution. In both panels the late time average of RGPE decaying turbulence is indicated by black lines. In the bottom panel, the spectra of thermal equilibria Abrikosov lattices obtained with RGLET are indicated by red lines. From light to dark red, the simulations with increasing temperature correspond to $T/T_c \approx 0.005$, $0.01$, $0.02$, and $0.05$. Several power laws are shown as references.}
    \label{fig:thermalspec}
\end{figure}

The thermal equilibrium with $T/T_c \approx 0.05$ qualitatively resembles the RGPE solution in Fig.~\ref{fig: realdensity_turbulence_decay}. To further compare both solutions, we look at their kinetic energy spectra (see Fig.~\ref{fig:thermalspec}). In the figure we show the compressible and incompressible kinetic energy spectra of the late time solutions of freely decaying turbulence with $\Omega = 1.2 \, U/L$ in RGPE, compared against spectra of the same energy components from thermal equilibria obtained with RGLET using the same $\Omega$ and different temperatures. The top panel compares the late time decay of turbulence in the condensate with the zero temperature Abrikosov lattice. The spectra are very different: the zero temperature Abrikosov lattice displays peaks at intermediate wave numbers, and flat spectra at larger wave numbers not observed in the long time decay of turbulence. Note that the incompressible kinetic energy spectrum of the RGPE freely decaying run peaks at $k=1$ and follows, for $k\gtrsim 20$, a power law compatible with $E_k^{(i)}(k) \sim k^{-1}$ spectrum\cite{AmetteEstrada2022}, i.e., a Vinen or ultraquantum-like scaling. The compressible kinetic energy spectrum for the same simulation has a slightly shallower scaling than $\sim k$ in the same range.

However, the thermal equilibria ($T>0$) obtained with RGLET share several similarities with the spectra of the freely decaying simulation (see the bottom panel of Fig.~\ref{fig:thermalspec}). For the incompressible kinetic energy, except for a pile up of energy for very large wavenumbers, the spectra are similar to $E_k^{(i)}(k)$ from RGPE. In other words, the peak at small wave numbers is associated to a large-scale flow generated by the wiggly Abrikosov lattice, and the $\sim k^{-1}$ spectrum at larger wave numbers results, as discussed in \cite{AmetteEstrada2022}, from the fluxless disordered array of vortices (note that the Abrikosov lattice at $T=0$ does not have this spectrum). For the compressible kinetic energy, the spectrum generates a quasi-2D thermalized state with scaling $\sim k$. For $k \gtrsim 20$ the amplitude of this spectrum in the RGLET run with $T/T_c \approx 0.05$ is closer to $E_k^{(c)}(k)$ in the RGPE run, but note also that all thermal equilibria have less compressible energy density at small wave numbers when compared with RGPE. This can be expected as the RGPE run starts with a random flow at intermediate scales that can emit sound waves at those scales, while RGLET runs are initialized with a Gaussian density profile at rest.

Thus, the free decay of rotating quantum turbulence seems to lead the system to a state that is partially thermalized and partially ordered, unlike classical behavior. As the flow decays, incompressible kinetic energy is mostly transferred towards large scales for $\Omega > \Omega_c$, where a large-scale flow compatible with the Abrikosov lattice is generated. However, a small fraction of that energy is transferred towards smaller scales, where through reconnection or a wave cascade it excites phonons. These phonons generate an effective thermal bath, shaking and wiggling the Abrikosov lattice. The system thus decays to a new equilibrium that has an Abrikosov lattice disordered by random fluctuations, and very similar to finite temperature rotating equilibria resulting from Canonical or Grand canonical ensembles.

\section{Conclusions}
\label{sec: conclusion}

We studied freely decaying rotating turbulence in BECs using the rotating Gross-Pitaevskii equation, with a special emphasis on how rotation affects the decay of turbulence. Global quantities were observed to present a sharp change in their decay or growth rates when the dimensionless controlling parameter $\Omega / \Omega_c \approx 1$, where $\Omega_c$ is the minimum rotation rate such that spontaneously creating one quantized vortex in the BEC to mimic the solid body rotation has less free energy than having none. For $\Omega / \Omega_c < 1$ the system evolves as in the non-rotating case, with a fast decay of incompressible kinetic energy, excitation of compressible modes, and vortex stretching associated to a three-dimensional direct cascade of incompressible kinetic energy. For $\Omega / \Omega_c > 1$, incompressible kinetic energy energy is mostly transferred inversely (i.e., to larger scales), and the excitation of compressible modes as well as vortex stretching are diminished. The flow becomes quasi-two-dimensional, with a tendency of the quantized vortices to align with the rotation axis. The behavior of the flow integral scale, as well as of the energy in the gravest Fourier modes, further confirm this behavior.

Rotating quantum turbulence decays for sufficiently long times into states that are partially ordered, displaying topological defects in a disordered arrangement reminiscent of an Abrikosov lattice. To compare against statistical equilibria, we presented a theoretical framework to generate finite temperature states of rotating BECs. Comparison of these thermalized states with the long time decay of rotating turbulence shows strong similarities in the spatial distribution of mass density, as well as in the spectral distribution of the incompressible and compressible kinetic energies. This confirms that the observed $\sim k^{-1}$ Vinen-like spectrum of incompressible kinetic energy can be the result of a random quasi-two-dimensional arrangement of vortices \cite{AmetteEstrada2022}, that the emission of phonons is mostly two-dimensional (as indicated by the $\sim k$ thermalized scaling of the compressible kinetic energy), and provides useful information on how rotating quantum turbulence decays into new equilibria.

\begin{acknowledgments}
JAE and PDM acknowledge financial support from UBACYT Grant No.~20020170100508BA and ANPCyT PICT Grant No.~2018-4298. MEB acknowledges support from the French Agence Nationale de la Recherche (ANR QUTE-HPC project No.~ANR-18-CE46-0013).
\end{acknowledgments}

\bibliography{ms}

\end{document}